\documentclass{sigchi}
\usepackage[T1]{fontenc}   %
\usepackage{amsmath}
\usepackage{amssymb}
\usepackage{balance}       %
\usepackage{graphics}      %
\usepackage{mathptmx}
\usepackage{color}
\usepackage[pdflang={en-US},pdftex]{hyperref}
\usepackage{booktabs}
\usepackage{times}
\usepackage{caption}
\usepackage{microtype}        %
\usepackage{ccicons}          %

\usepackage{xspace}
\usepackage{enumitem}
\usepackage{float}
\usepackage{multirow}

\DeclareRobustCommand{\etal}{\textit{et al.}\xspace}
\newcommand*{\modern}{\fontfamily{cmss}\selectfont}

\DeclareRobustCommand{\etal}{\textit{et al.}\xspace}\DeclareRobustCommand{\surl}[1]{\emph{\urlstyle{same}\url{#1}}\xspace}

\DeclareRobustCommand{\bp}{\emph{backward projection}\xspace}
\DeclareRobustCommand{\Bp}{\emph{Backward projection}\xspace}
\DeclareRobustCommand{\fp}{\emph{forward projection}\xspace}
\DeclareRobustCommand{\Fp}{\emph{Forward projection}\xspace}

\DeclareRobustCommand{\fm}{\emph{feasibility map}\xspace}
\DeclareRobustCommand{\pl}{\emph{prolines}\xspace}
\DeclareRobustCommand{\Pl}{\emph{Prolines}\xspace}

\DeclareRobustCommand{\surl}[1]{\emph{\urlstyle{same}\url{#1}}\xspace}
\definecolor{orange}{rgb}{1, 0.49, 0.05}
\definecolor{gray}{rgb}{0.49, 0.49, 0.49}

\setcopyright{none}
\doi{}
\isbn{}

\def\plaintitle{Exploring Dimensionality Reductions with Forward and Backward Projections}

\def\emptyauthor{}
\def\plainkeywords{
  Dimensionality reduction; 
  interaction; 
  bidirectional binding; 
  visual embedding; 
  forward projection; 
  backward projection; 
  PCA; 
  autoencoder; 
  prolines; 
  feasibility map; 
  what-if analysis; 
  Praxis.}

\makeatletter
\def\url@leostyle{%
  \@ifundefined{selectfont}{
    \def\UrlFont{\sf}
  }{
    \def\UrlFont{\small\bf\ttfamily}
  }}
\makeatother
\def\pprw{8.5in}
\def\pprh{11in}

\definecolor{linkColor}{RGB}{6,125,233}
\hypersetup{%
  pdftitle={\plaintitle},
  pdfauthor={\emptyauthor},
  pdfkeywords={\plainkeywords},
  pdfdisplaydoctitle=true, %
  bookmarksnumbered,
  pdfstartview={FitH},
  colorlinks,
  citecolor=black,
  filecolor=black,
  linkcolor=black,
  urlcolor=linkColor,
  breaklinks=true,
  hypertexnames=false
}

\begin{document}
\title{\plaintitle}

\numberofauthors{2}
\author{%
  \alignauthor{Marco Cavallo\\
    \affaddr{IBM Research}\\
    \email{mcavall@us.ibm.research}}\\
  \alignauthor{\c{C}a\u{g}atay Demiralp\\
    \affaddr{IBM Research}\\
    \email{cagatay.demiralp@us.ibm.research}}\\
}

\maketitle

\begin{abstract}
Dimensionality reduction is a common method for analyzing and visualizing
high-dimensional data across domains. Dimensionality-reduction algorithms
involve complex optimizations and the reduced dimensions computed by these
algorithms generally lack clear relation to the initial data dimensions.
Therefore, interpreting and reasoning about dimensionality reductions can be
difficult. In this work, we introduce two interaction techniques, \fp and \bp,
for reasoning dynamically about scatter plots of dimensionally reduced data. We
also contribute two related visualization techniques, \pl and \fm, to
facilitate and enrich the effective use of the proposed interactions, which we
integrate in a new tool called \textit{Praxis}. To evaluate our techniques, we
first analyze their time and accuracy performance across varying sample and
dimension sizes.  We then conduct a user study in which twelve data scientists
use \textit{Praxis} so as to assess the usefulness of the techniques in
performing exploratory data analysis tasks. Results suggest that our visual
interactions are intuitive and effective for exploring dimensionality
reductions and generating hypotheses about the underlying data.
\end{abstract}


\section{Introduction}

\begin{figure*}[th]
 \centering
 \includegraphics[width=0.45\linewidth]{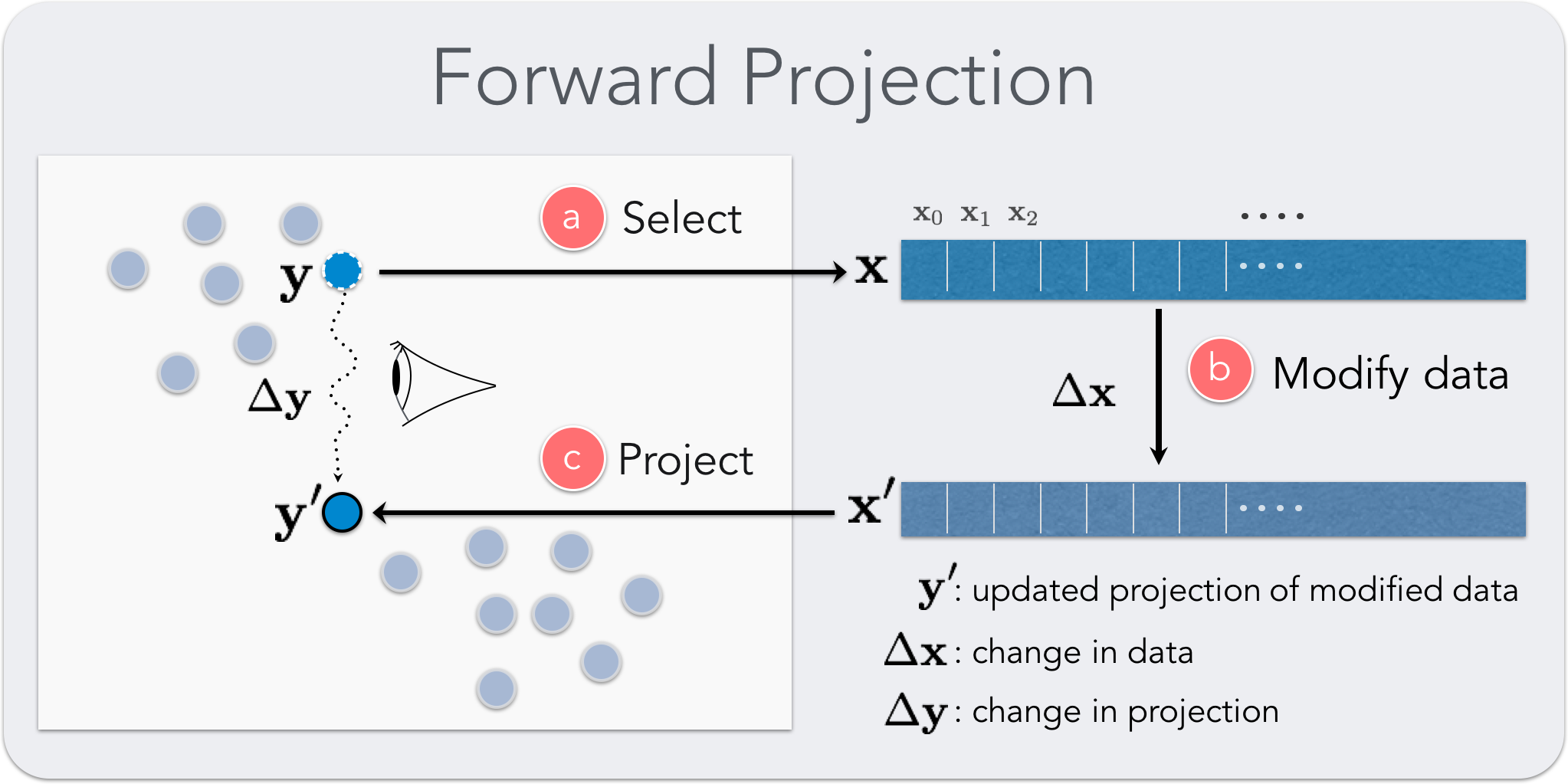}
 \enskip
 \includegraphics[width=0.45\linewidth]{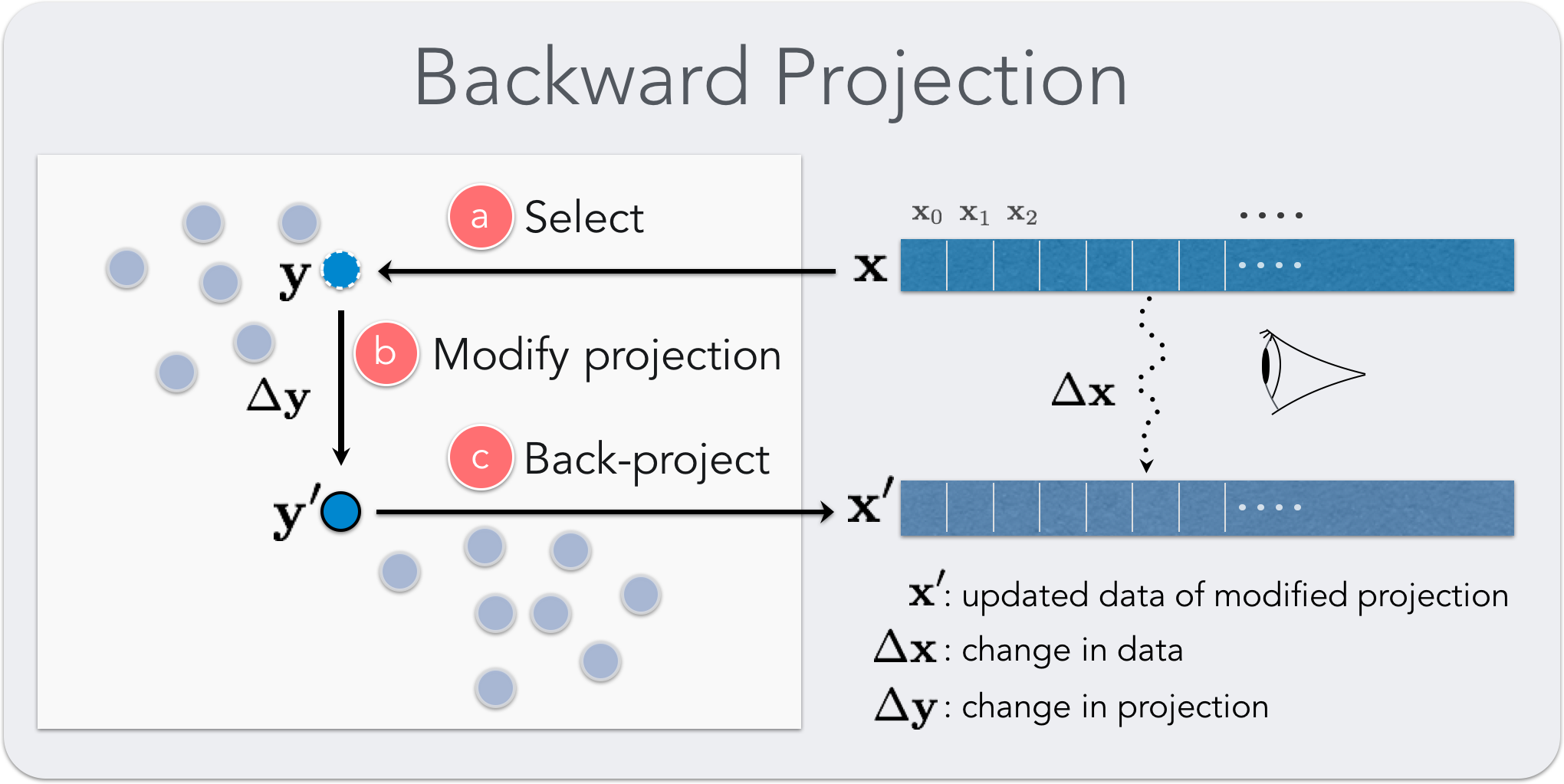}
  \caption{\normalfont  (Left) Forward
    projection enables users to: (a) select any data point $\mathbf{x}$, (b)
    interactively change its high-dimensional feature values, and (c) observe
    the change $\Delta \mathbf{y}$ in the point's two-dimensional projection.
    (Right) Backward projection enables users to: (a) select any node
    corresponding to the two-dimensional projection of a data point
    $\mathbf{x}$, (b) move the node arbitrarily in the plane, and (c) observe the chage 
    $\Delta \mathbf{x}$ in the point's high-dimensional feature values. \label{fig:teaser}
  } 
 \end{figure*}

Dimensionality reduction (DR) is an effective technique for analyzing and
visualizing high-dimensional datasets across domains, from sciences to
engineering. Dimensionality-reduction algorithms such as principal component
analysis (PCA) and multidimensional scaling (MDS) automatically reduce the
number of dimensions in data while maximally preserving structures, typically
quantified as similarities, correlations or distances between data points. This
makes visualization of the data possible using conventional spatial techniques.
For example, analysts generally use scatter plots to visualize the data after
reducing the number of dimensions to two, encoding the reduced dimensions in a
two-dimensional position. 

\noindent{\bf DR Challenges:} Most DR (also called manifold learning or
distance embedding) algorithms  are driven by complex numerical optimizations.
Dimensions derived by these methods generally lack clear, easy-to-interpret
mappings to the original data dimensions, forcing users to treat DR methods as
black boxes.  In particular, data analysts with limited experience in DR have
difficulty in interpreting the meaning of the projection axes and the position
of scatter plot nodes~\cite{Sedlmair_2013, Brehmer_2014}. \textit{What do the
axes mean?} is probably users' most frequent question when looking at scatter
plots in which points (nodes) correspond to dimensionally reduced data.  Most
scatter-plot visualizations of dimensionally reduced data are viewed as static
images. One reason is that tools for computing and plotting these
visualizations, such as Matlab and R, provide limited interactive exploration
functionalities. Another reason is that few interaction and visualization
techniques that go beyond brushing-and-linking or cluster-based coloring to 
allow dynamic reasoning with these visualizations.

\noindent{\bf Enriching User Experience with DRs:} In this paper, we introduce
two interactions, \fp and \bp (Figure~\ref{fig:teaser}), to help analysts  explore
and reason about scatter plot representations of dimensionally reduced data,
facilitating a dynamic what-if analysis. We contribute two related
visualization techniques, \pl and \fm, to facilitate the effective use of the
proposed interactions. We also introduce Praxis, a new interactive DR exploration 
tool implementing our interaction and visualization techniques for data analysis. 

Our techniques enable users to interactively explore: 1) the most
important features that determine the vertical and horizontal axes of
projections, 2) how changing feature values (dimensions) of a data point
changes the point's projected location (two-dimensional representation) and 3)
how changing the projected position of a data point changes the high-dimensional
values of that point. We demonstrate our techniques first using a
PCA-based linear DR and then a nonlinear, deep
autoencoder-based ~\cite{hinton2006reducing} DR.

 We assess the computational effectiveness of our methods by analyzing their
 time and accuracy performances under varying sample and dimension sizes.  We
 then conduct a user study in which twelve data scientists performed
 exploratory data analysis tasks using Praxis. The results suggest that our
 visual interactions are scalable and intuitive and can be effective for
 exploring dimensionality reductions and generating hypotheses about the
 underlying data.

 We also observe that our techniques belong to a class of interactions that
 bidirectionally couple the data and its visual representation: \emph{dynamic
 visualization interactions}~\cite{Victor_2013}. We look at dynamic
 visualization interactions under the visual embedding
 model~\cite{Demiralp_2014} and discuss the properties of effective
 interactions that the model suggests.   
 
\section{Related Work}
Our work is related to prior efforts in understanding and 
improving user experience with dimensionality reductions. 

\subsection{Direct Manipulation in DR}
Direct manipulation has a long history in human-computer
interaction~\cite{sutherland1964sketchpad,kay1977personal,borning1981programming}
and visualization research (e.g.~\cite{shneiderman1982direct}).  Direct
manipulation techniques aim to improve user engagement by minimizing the
\textit{perceived} distance between the interaction source and the target
object~\cite{hutchins1985direct}.

Developing direct manipulation interactions to guide DR formation and modify
the underlying data is a focus of prior research~\cite{Buja_2008,Endert_2012,
Gleicher_2013,Jeong_2009,Johansson_2009,Williams_2004}.  For example,
X/GGvis~\cite{Buja_2008} supports changing the weights of dissimilarities input
to the MDS stress function along with the coordinates of the embedded  points
to guide the projection process. Similarly, iPCA~\cite{Jeong_2009} enables
users to interactively modify the weights of data dimensions in computing
projections. Endert \etal \cite{Endert_2011} apply similar ideas to additional
dimensionality-reduction methods while incorporating user feedback through
spatial interactions in which users can express their intent by dragging points
in the plane.

Our work is closely related to earlier approaches using direct manipulation to
modify data in DR visualizations~\cite{Jeong_2009,
viau2010flowvizmenu,Schreck_2009,crnovrsanin2009proximity,mamani2013user}.
Like our \fp and unconstrained \bp techniques, iPCA enables interactive forward
and backward projections for PCA-based DRs. However, iPCA recomputes full PCA
for each forward and backward projection, and these can suffer from jitter and
scalability issues, as noted in \cite{Jeong_2009}.  Using out-of-sample
extrapolation, \fp avoids re-running dimensionality-reduction algorithms. From
the visualization point of view, this is not just a computational convenience,
but also has perceptual and cognitive advantages, such as preserving the
constancy of scatter-plot representations.  For example, re-running (training)
a dimensionality reduction algorithm with a new data sample added can
significantly alter a two-dimensional scatter plot of the dimensionally reduced
data, even though all the original inter-data point similarities may remain
unchanged. In contrast to iPCA, we also enable users to interactively define
constraints on feature values and perform constrained \bp. 
  
We refer readers to a recent survey~\cite{sacha2017visual} for a detailed
discussion of prior research on visual interaction with dimensionality
reduction. 

\subsection{Visualization in DR Scatter Plots}

Prior work introduces various visualizations in planar scatter plots of DRs, in
order to improve the user experience by communicating projection
errors~\cite{Chuang_2012,Stahnke_2016,Aupetit_2007,Lespinats_2010}, change in
dimensionality projection positions~\cite{Jeong_2009}, data properties and
clustering results~\cite{Stahnke_2016,clustrophile:idea16}, and contributions
of original data dimensions in reduced dimensions~\cite{gabriel1971biplot}.
Low-dimensional projections are generally lossy  representations of the
original data relations: therefore, it is useful to convey both overall and
per-point dimensionality-reduction errors to users when desired.  Researchers
visualized  errors in DR scatter plots using Voronoi
diagrams~\cite{Aupetit_2007,Lespinats_2010} and corrected (undistorted) the
errors by adjusting the projection layout with respect to the examined
point~\cite{Chuang_2012,Stahnke_2016}.

Biplot was introduced~\cite{gabriel1971biplot} to visualize the magnitude and
sign of a data attribute's contribution to the first two or three principal
components as line vectors in PCA. \Pl reduce to biplots when PCA is used for
dimensionality reduction.  Our \textit{proline} construction algorithm is
general and reflects the underlying out-of-sample extension method used. On the
other hand, biplots are based on singular-value decomposition and always use
PCA forward projection, regardless  of the actual DR used. Additionally, \pl
differ from biplots in being interactive visual objects beyond static vectors
and are decorated to communicate distributional characteristics of the
underlying data point. 

Stahnke \etal ~\cite{Stahnke_2016} use a grayscale map to visualize how a
single attribute value changes between data points in DR scatter
plots.  We use \fm, a grayscale map, to visualize the feasible regions in the
constrained \bp interaction.

\subsection{Out-of-sample Extension and Back Projection for DR}

We compute forward projections using out-of-sample extension (or
extrapolation)~\cite{Maaten_2009}. Out-of-sample extension is the projection of
a new data point into an existing DR (e.g. learned manifold model) using only
the properties of the already computed DR. It is conceptually equivalent to
testing a trained machine-learning model with  data that was not part of the
training set. For linear DR  methods, out-of-sample extension is often
performed by applying the learned linear  transformation to the new data point.
For autoencoders, the trained network  defines the transformation from the
high-dimensional to low-dimensional data representation~\cite{Bengio_2004}. 
  
Back or backward projection maps a low-dimensional data point back into the
original high-dimensional data space. For linear DRs, back projection is
typically done by applying the inverse of the learned linear DR mapping. For
nonlinear DRs, earlier research proposed DR-specific backward-projection
techniques. For example, iLAMP~\cite{dos2012ilamp} introduces a back-projection
method for LAMP~\cite{joia2011lamp} using local neighborhoods and demonstrates
its viability over synthetic datasets~\cite{dos2012ilamp}.  Researchers also
investigated general backward projection methods   using radial basis
functions~\cite{monnig2014inverting,amorim2015facing}, treating backward
projection as an interpolation problem. 

Autoencoders~\cite{hinton2006reducing}, neural-network-based DR models, are a
promising approach to computing backward projections.  An autoencoder model
with multiple hidden layers can learn a nonlinear dimensionality reduction
function (encoding) as well as the corresponding backward projection
(decoding) as part of the DR process. Inverting DRs is, however, an ill-posed
problem. In addition to augmenting what-if analysis, the ability to define
constraints over a back projection can also ease the computational burden.
Praxis also enables users to interactively set equality and boundary
constraints  over back projections through an intuitive interface. 

We presented initial versions of \fp, \bp, and \pl earlier as part of
Clustrophile, an exploratory visual clustering analysis
tool~\cite{clustrophile:idea16}. We give here a focused discussion of our
revised interaction and visualization techniques, introduce Praxis, a new
visualization tool that implements our techniques for exploratory data analysis
using DR, and provide a thorough computational and user-performance evaluation.
The current work also introduces \fm, a new visualization technique to
facilitate \bp interactions.      

\section{Interacting with Linear Dimensionality Reductions}
We demonstrate our methods first on principal component analysis (PCA), one of
the most frequently used linear dimensionality-reduction techniques; note that
the discussion here applies as well to other linear dimensionality-reduction
methods. PCA computes (learns) a linear orthogonal transformation of the
empirically centered data into a new coordinate frame in which the axes
represent maximal variability. The orthogonal axes of the new coordinate frame
are called principal components. 

To reduce the number of dimensions to two, for example,
we project the centered data matrix, rows of which correspond to data
samples and columns to features (dimensions), onto the first two principal
components, $\mathbf{e_0}$ and $\mathbf{e_1}$. 
Details of PCA along with its many formulations and interpretations can be
found in standard textbooks on machine learning or data mining (e.g.,
\cite{Bishop_2006,Hastie_2005}).

\subsection{Forward Projection}

\begin{figure}[t]
\centering
\includegraphics[width=\linewidth]{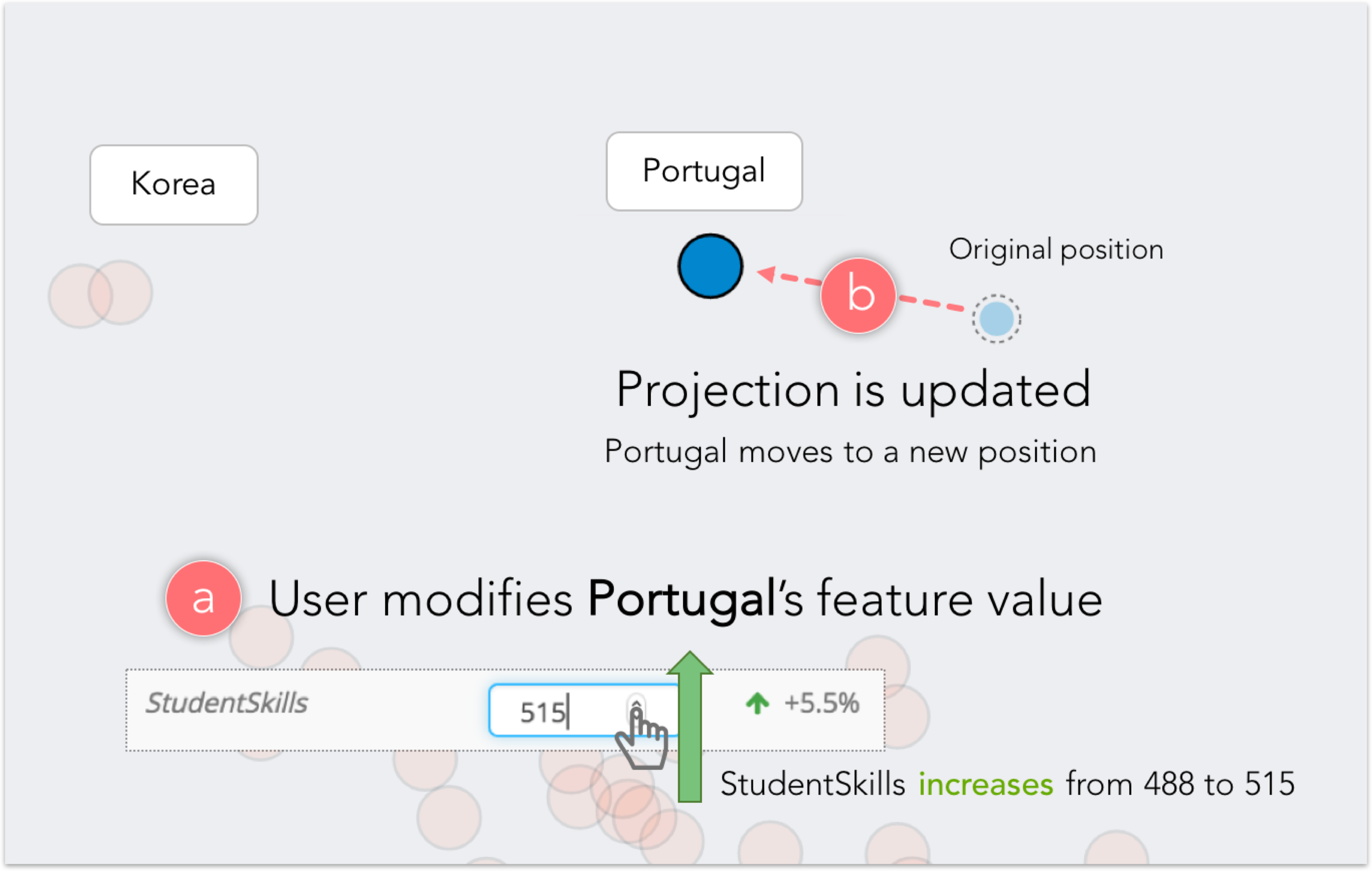}
\caption{\normalfont Through forward projection, a user can quickly explore how much the 
  difference in the {\modern StudentSkills} index value explains the planar projection 
  difference between Portugal (blue node) and Korea.\label{fig:fpinaction}} 
\end{figure}

\Fp enables users to interactively change the feature values  of a data point
$\mathbf{x}$ and observe how these hypothesized changes in data modify the
current projected location $\mathbf{y}$ (Figure~\ref{fig:fpinaction}). This is
useful  because understanding the importance and sensitivity of features
(dimensions) is a key goal in exploratory data analysis.  In the case 
of PCA, we obtain the two-dimensional position change vector
$\Delta\mathbf{y}$ by projecting the data change vector $\mathbf{x^\prime}$
onto the principal components: $\Delta\mathbf{y} = \Delta
\mathbf{x}\;\mathbf{E}$, where $\mathbf{E} = \begin{bmatrix}\mathbf{e_0} &
  \mathbf{e_1} \end{bmatrix}$. 

\begin{figure}[t]
\centering
\includegraphics[width=\linewidth]{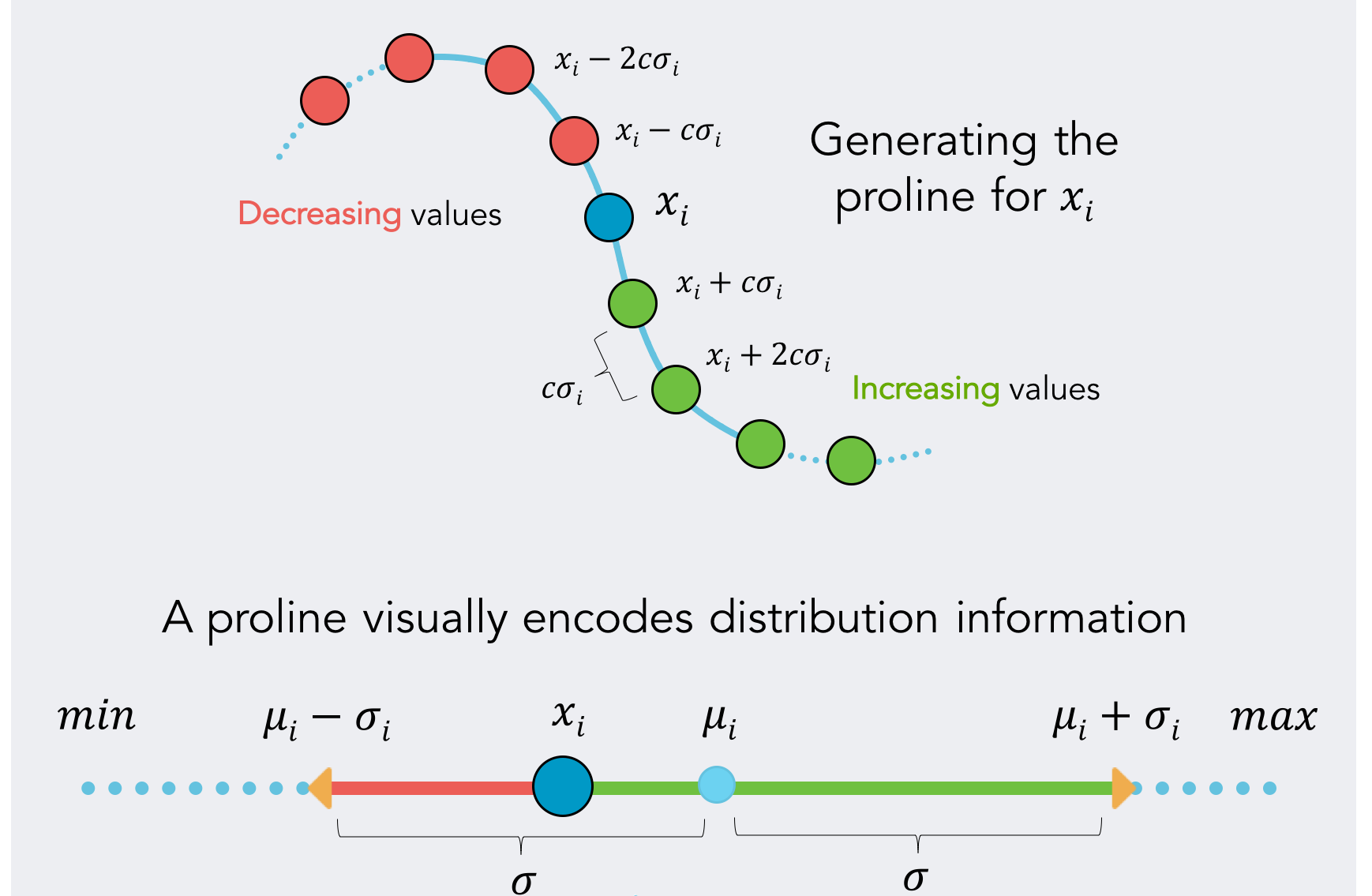}

\caption{ \normalfont Proline construction. For a given dimension (feature) ${x}_i$ of
  a point $\mathbf{x}$ in a dataset  explored, we construct a proline by
  connecting the forward projections of data points regularly sampled from a
  range of $\mathbf{x}$ values, where all features are fixed but ${x}_i$
  varies. A proline also encodes the forward projections for the ${x}_i$ values
  in $\left[\mu_i-\sigma_i, \mu_i+\sigma_i\right]$ with thick green and red
  line segments, providing a basic directional statistical context. $\mu_i$ is
  the mean of the $i$th dimension in the dataset, the green segment represents
  forward projections for ${x}_i$ values in $\left[{x}_i,
  \mu_i+\sigma_i\right]$ and the red segment for ${x}_i$ values in
  $\left[\mu_i-\sigma_i, {x}_i\right]$.\label{fig:pl}}
\end{figure}

\begin{figure}[t]
\centering
\includegraphics[width=\linewidth]{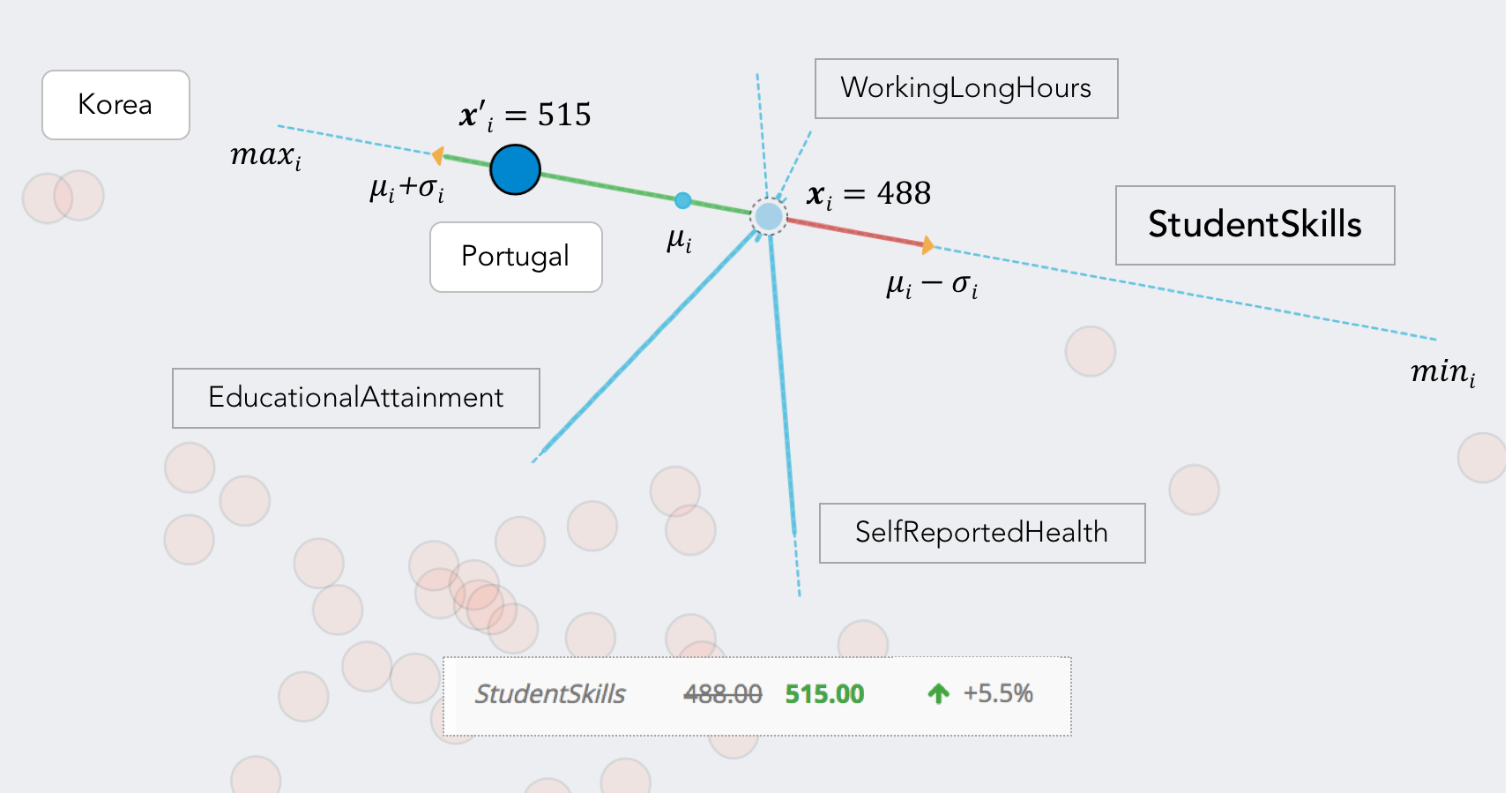}
\caption{
  \normalfont Forward projection with prolines. StudentSkills is 
  revealed as key feature differentiating
  Portugal from Korea. Observe that a value of 515 for StudenSkills would 
  be reasonable with respect to the feature
  distribution ($\mu_i < 515 < \mu_i+\sigma_i$), but not enough to make
  Portugal close to Korea in the projection plane.  By visually comparing the
  lengths (variability) of different proline paths, the user can easily
  recognize which dimensions contribute most to determining the position of
  points in the dimensionally reduced space. For instance, a change in the
  feature value of WorkingLongHours (the shortest proline) would produce only a
  very small change in the projection.\label{fig:fp_prolines}} 
\end{figure}

\subsection{Prolines: Visualizing Forward Projections}
It is desirable to see in advance what forward projection paths  look like for
each feature. Users can then start inspecting the dimensions that look
interesting or important. 

\Pl visualize forward projection paths using a linear range of possible values
for each feature and data point (Figures~\ref{fig:pl}).  Let $\mathbf{x}_i$ be
the value of the $i$th feature for the data point $\mathbf{x}$. We first
compute the mean $\mu_i$, standard deviation $\sigma_i$, minimum $min_i$ and
maximum $max_i$ values for the feature in the dataset  and devise a range $ I =
\left[min_i,  max_i\right]$. We then iterate over the range with step size
$c\sigma_i$, compute the forward projections as discussed above, and then
connect them as a path. The constant $c$ controls the step size with which we
iterate over the range and is set to $c=\sigma_i/8$ for the examples shown
here.  

In addition to providing an advance snapshot of forward projections, \pl can be
used to provide summary information conveying the relationship between the
feature distribution and  the  projection space.  We display along  each
\textit{proline} a small light-blue circle indicating the position that the
data point would  assume if it had a feature value corresponding to the mean of
its distribution; similarly, we display two small arrows indicating a variation
of one standard deviation ($\sigma_i$) from the mean ($\mu_i$). The segment
identified by the range $\left[\mu_i - \sigma_i, \mathbf{x}_i +
\sigma_i\right]$ is highlighted and further divided into two segments. The
green segment shows the positions that the data point would assume by
increasing its feature value; the red one indicates a decreasing value. This
enables users to infer the relationship  between the feature space and the
direction of change in the projection space.

\begin{figure}[t]
\centering
\includegraphics[width=\linewidth]{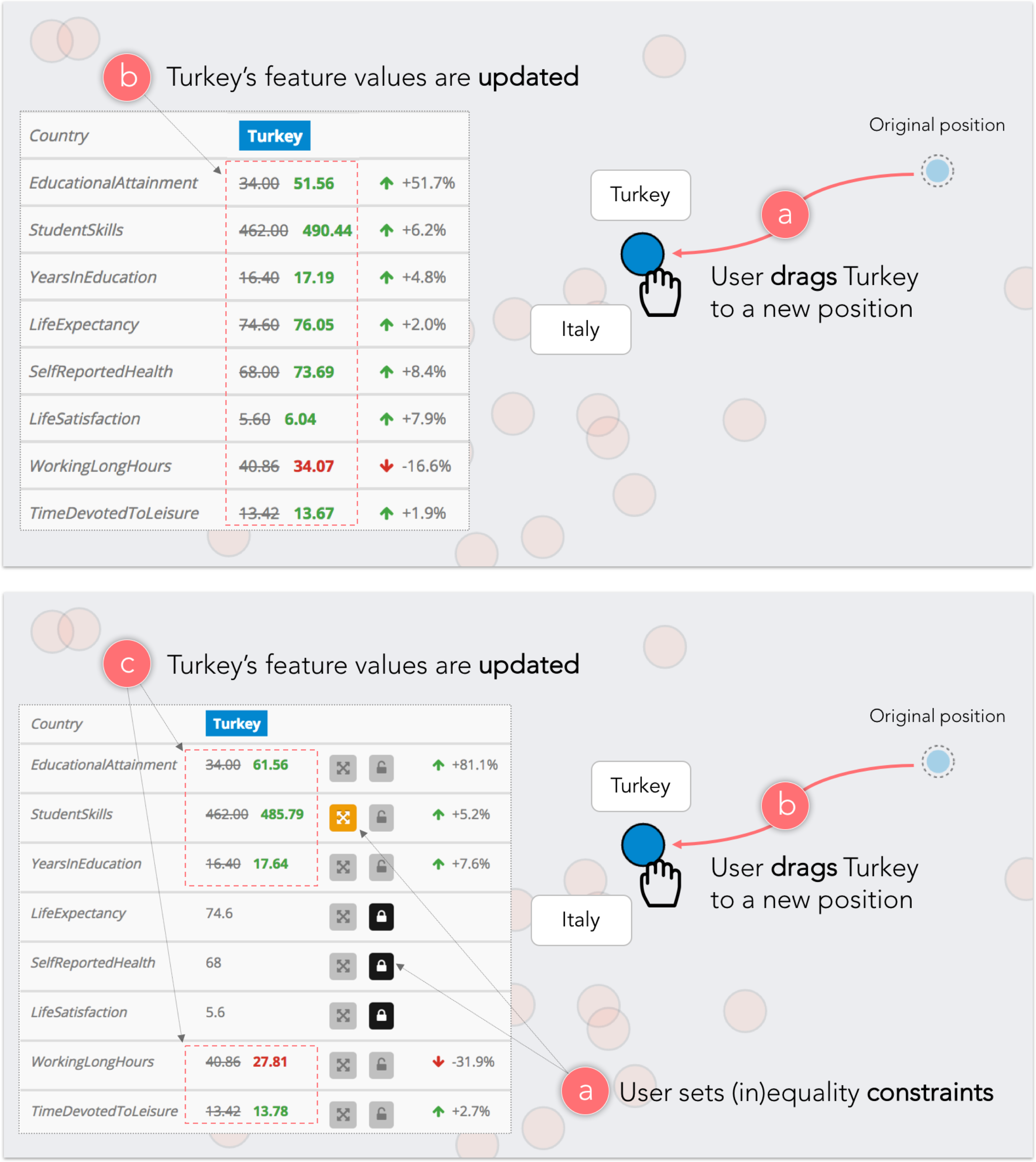}
\caption{\normalfont (Top) Unconstrained backward projection usage: 
  Curious about the projection difference between
 Turkey and Italy, similar countries in some respects, the user moves the node
associated with Turkey (blue circle) towards Italy (a). The feature  values of
Turkey are automatically updated (b) to satisfy the new projected position as
the node is moved. (Bottom) Constrained backward projection usage. Considering 
that the features {\modern LifeExpectancy},
{\modern SelfReportedHealth} and {\modern LifeSatisfaction} are unmodifiable, 
the user puts a lock (equality constraint) on their values. Through a dedicated
interface (Figure~\ref{fig:praxis}d) the user also sets the upper bound for the
feature {\modern StudentSkills} to 490 (inequality constraint). When performing
\bp, the feature values of Turkey are updated in order to respect the
user-defined constraints (c). \label{fig:bpinaction}}

\end{figure}

\subsection{Backward Projection}

\Bp as an interaction technique is a natural complement of \fp. Consider the
following scenario: a user looks at a projection and, seeing a cluster of
points and a single point projected far from this group, asks what changes in
the feature values of the outlier point would bring the outlier near
the cluster. Now, the user can play with  different dimensions using \fp to
move the current projection of the outlier point near the cluster. It would be
more natural, however, to move the point directly and observe the change. 

The formulation of \bp is the same as that of \fp:  $\Delta \mathbf{y} = \Delta
\mathbf{x}\;\mathbf{E}$.  In this case, however, $\Delta \mathbf{x}$ is unknown
and we need to solve the equation.  

As formulated, the problem is underdetermined and, in general, there can be
an infinite number of data points (feature values) that project to the same planar
position. Therefore, we propose both unconstrained and constrained backward
projections, for which users can define equality as well as inequality constraints.  

In the case of unconstrained backward projection, we find $\Delta \mathbf{x}$ 
by solving a regularized least-squares optimization problem: 

\begin{equation*}
  \begin{aligned}
    & \underset{\Delta\mathbf{x}}{\text{minimize}}
    & & \|{\Delta\mathbf{x}}\|^2 \\
    & \text{subject to}
    & & \Delta\mathbf{x}\;\mathbf{E} = \Delta\mathbf{y}
  \end{aligned}
\end{equation*}
Note that this is equivalent to setting $\Delta \mathbf{x}=
\Delta\mathbf{y}\;\mathbf{E}^T$.

For constrained backward projection, we find $\Delta \mathbf{x}$ 
by solving the following quadratic optimization problem:   

\begin{equation*}
  \begin{aligned}
    & \underset{ \Delta\mathbf{x}}{\text{minimize}} & & \|{\Delta\mathbf{x}}\;\mathbf{E} - \Delta\mathbf{y}\|^2 \\
    & \text{subject to} & & \mathbf{C}\Delta\mathbf{x}  = \mathbf{d}\\
    & & & \mathbf{lb} \leq \Delta\mathbf{x} \leq \mathbf{ub} 
  \end{aligned}
\end{equation*}
$\mathbf{C}$ is the design matrix of equality constraints, $\mathbf{d}$ is the 
constant vector of equalities,  and $\mathbf{lb}$ and $\mathbf{ub}$ are the vectors 
of lower and upper boundary constraints.

\subsection{Guiding Backward Projections}

\begin{figure}[t]
\centering
\includegraphics[width=\linewidth]{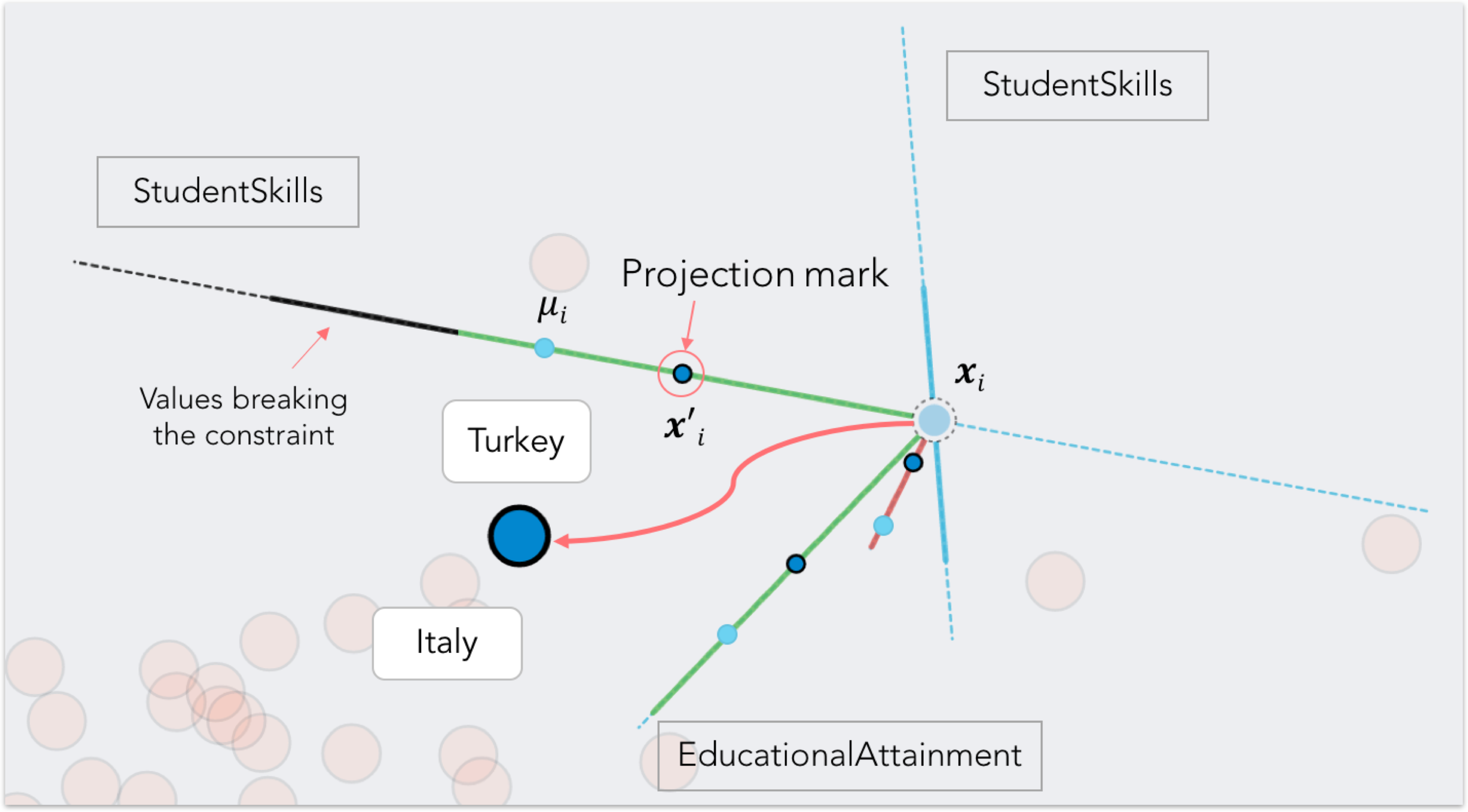}
\caption{\normalfont  With the addition of
\textit{projection marks}, \pl can be used as guides while the user performs
\bp. The figure shows the same constrained \bp example as
Figure~\ref{fig:bpinaction}: while the user drags the projected data point,
prolines indicate with a green or red color if their corresponding feature
value is increasing or decreasing. Projection marks, represented as little blue
circles, move along each proline as the user performs \bp and indicate the
current feature values. This is particularly useful for checking the position
of each value with respect to its feature distribution. In the case of
inequality constraints, values that do not satisfy a constraint generate a
black proline. \label{fig:bp_prolines}}
\end{figure}

\noindent{\bf Projection Marks:} It is important to note that, since more than
one data point in the multidimensional space can project to the same position,
forward and backward projections may not always correspond. For this reason, we
add to our prolines visualization a set of \textit{projection marks}
(Figure~\ref{fig:bp_prolines}) dynamically indicating the current value for
each feature while the user performs \bp. At the same time, dragging a data
point highlights the green or red segment of each proline based on the increase
or decrease of each feature, showing which dimensions correlate to each other.
By combining forward projection paths to \bp, the user can infer how fast each
value is changing in relation to its feature distribution.

\begin{figure}[t]
\centering
\includegraphics[width=\linewidth]{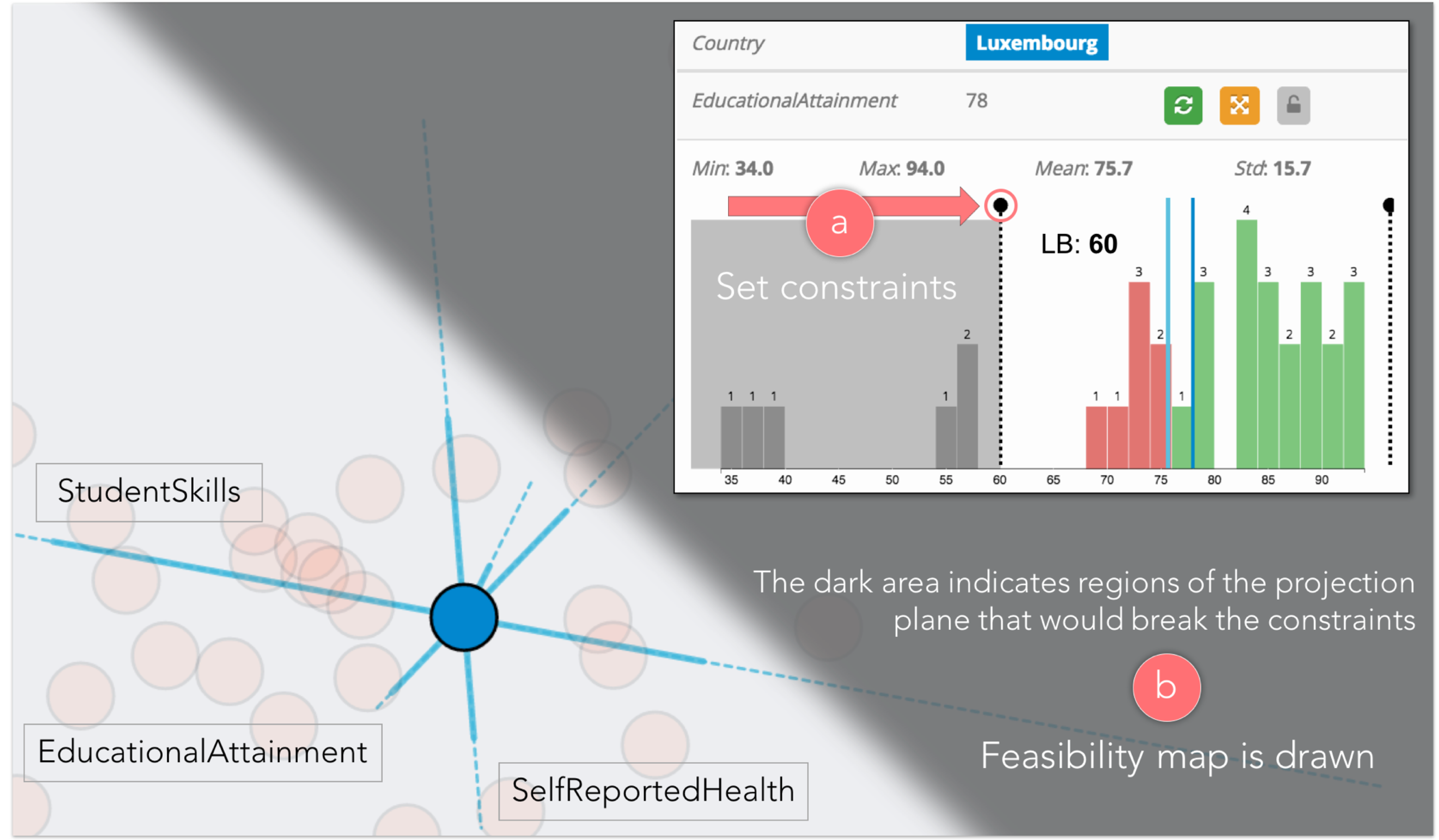}
\caption{\normalfont Feasibility map. The feasibility map samples the
  projection plane through backward projection and shows which regions are not
  admissible based on a set of user-defined constraints.  Here (a) a user
  defines a lower bound for the {\modern
  EducationalAttainment} value through the interface provided (described in
  Section~\ref{sec:praxis}) and (b) a dark area is drawn onto the projection
  plane, indicating that moving the projected point (Greece) in that 
  region would break the constraint.\label{fig:fm}}
\end{figure}

\noindent {\bf Feasibility Maps:} Constrained backward projection enables users
to semantically regulate the mapping into unprojected high-dimensional data
space. For example we don't expect an Age field to be negative or bigger than
150, even though such a value can can constitute a more optimal solution in an
unconstrained backward projection scenario. 

We propose the \fm visualization as a way to quickly see the feasible space
determined by  a given set of constraints.  Instead of manually checking  if a
position in the projection plane satisfies the desired range of values
(considering both equality and inequality constraints), it is desirable to know
in advance which regions of the plane correspond to admissible solutions.  In
this sense, \fm is a conceptual generalization of \pl to the constrained backward projection
interaction. 

To generate a \fm, we sample the projection plane on a regular grid and evaluate
the feasibility at each grid point based on the constraints imposed by the user,
obtaining a binary mask over the projection plane.  We render this binary mask
over the projection as an interpolated grayscale heatmap, where darker areas
indicate infeasible planar regions (Figure~\ref{fig:fm}). With accuracy
determined by the number of \bp samples, the user can see which areas a data
point can assume in the projection plane without breaking the constraints.
Note that, when dealing with linear dimensionality reduction techniques, the
\fm originates boundaries that are orthogonal to the prolines of the
constrained features. Nevertheless, generating the \fm by sampling the
projection plane has the advantage of being independent of the dimensionality
technique used.

\begin{figure}
\centering
\includegraphics[width=\linewidth]{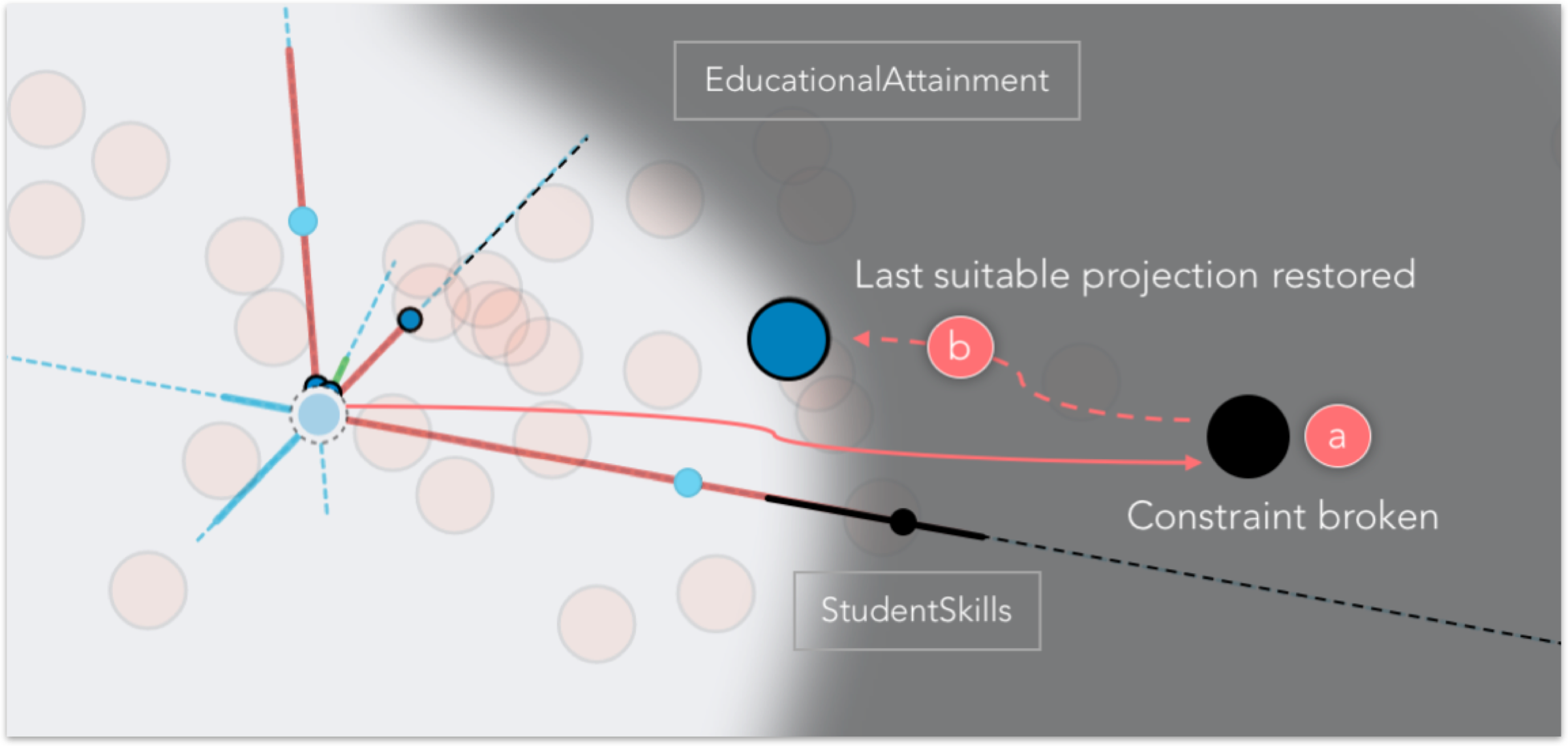}
\caption{\normalfont Visualizing broken constraints. If a projected point is dragged
  through backward projection onto an infeasible region (a), its color and the
  color of the projection marks associated with broken constraints turn black. 
  When the point is released, its position is restored to the last admissible 
  value computed through backward projection (b).
\label{fig:fminteraction}}
\end{figure}

In \bp, if a data point is dragged to a position that does not satisfy a
constraint, its color and the color of its corresponding projection marks turn
to black. If the user drops the data point in an infeasible position, the point
is automatically moved through animation back to the last feasible position 
to which it was dragged (Figure~\ref{fig:fminteraction}).

\section{Praxis}\label{sec:praxis}

\begin{figure*} \centering
  \includegraphics[width=0.9\linewidth]{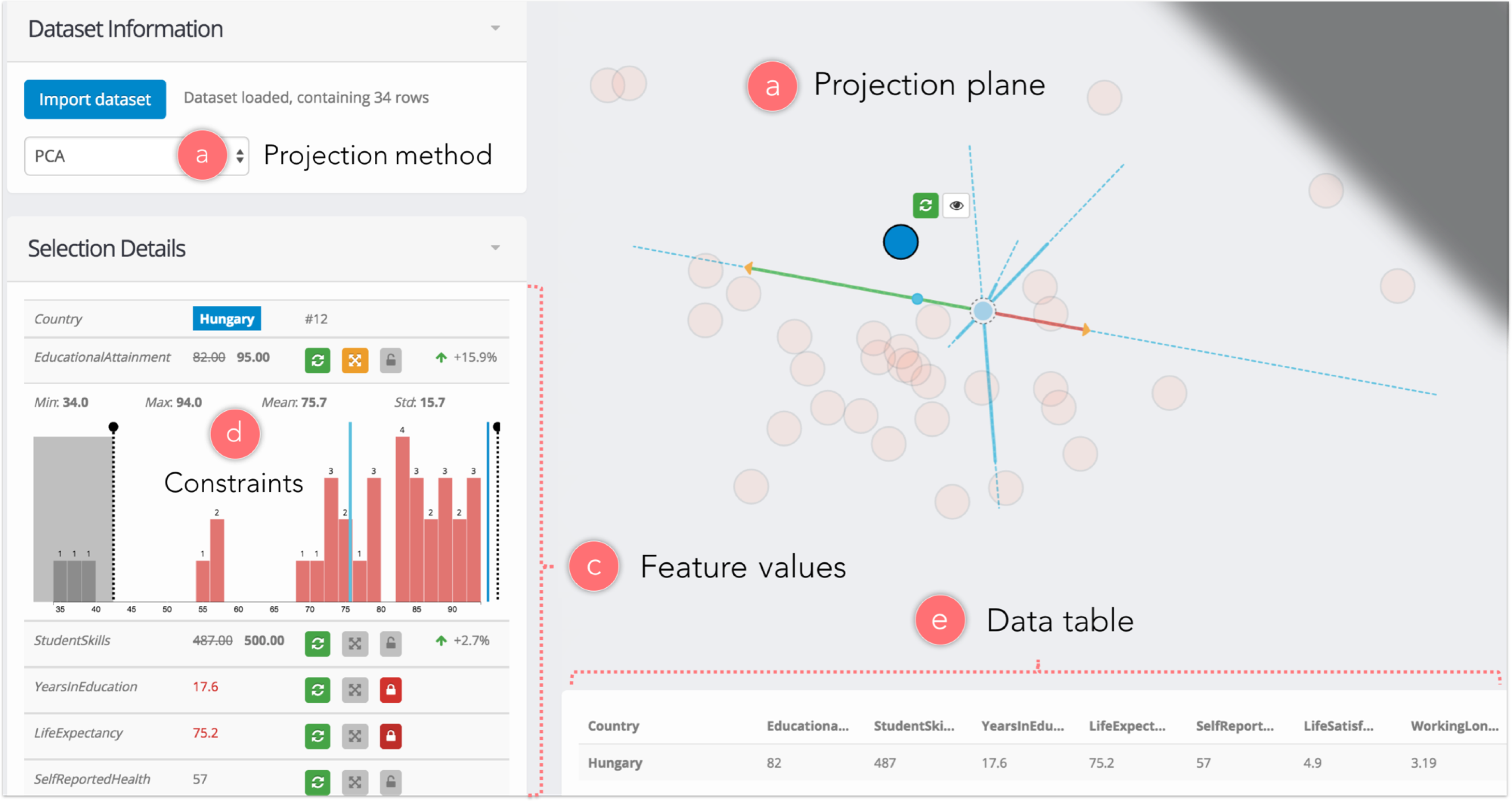}
  \caption{\normalfont Praxis interface.  Praxis is an interactive tool for
    DRs, integrating our projection interactions along with the related
    visualizations.  After importing a dataset and choosing a projection method
    (a), a scatter plot is displayed using the reduced two dimensions (b). When
    a point is selected, its feature values can be seen and modified from a
    table panel (c) that also allows entering constraints for each feature by
    double-clicking on a specific row of the table (d). A data table listing
    all rows in the dataset is also included (e). \label{fig:praxis}}
  \end{figure*}

To study the usage of our interaction and visualization techniques, we
introduce Praxis,  a new interactive tool integrating them. Through a data
panel (Figure~\ref{fig:praxis}b), users can load a dataset in CSV format and
visualize its PCA projection as a  scatter plot (Figure~\ref{fig:praxis}a),
using the first two principal components as axes of the projection plane. 
 
Results of forward and backward projection, along with the two visualizations
prolines and feasibility map, are displayed in the projection plane. The id
(name) of a data point is shown on mouse hover, while clicking performs
selection, showing its feature values in a dedicated sidebar panel
(Figure~\ref{fig:praxis}c).  In particular, the \textit{Selection Details}
panel is used for performing forward projection (clicking on a dimension makes
its value modifiable) and for inspecting changes in feature values when
backward projection is used. Three buttons enable (respectively for each
feature) 1) reset it to its original value, 2) enable/disable inequality
constraints and 3) lock its value to a specific number (equality constraint).

Double-clicking the row associated with a feature displays a histogram
representing its distribution below the selected row, showing some basic
statistics (Figure~\ref{fig:praxis}d). The current value of the feature is
represented by a blue line and a cyan line indicates the distribution mean.
Bins of the histogram are colored similarly to \pl: green for increasing
values, red for decreasing values with respect to the original feature value.
Dragging one of the two black handles lets the user set or unset lower and
upper bounds for a feature distribution, thus defining a set of constraints for
a specific data point.

Finally, selecting a data point in the projection plane displays two buttons
that respectively enable 1) resetting its feature values (and position) to
their original value and 2) showing a tooltip  on top of its $k$ currently
nearest neighbors, in order to facilitate reasoning about the similarity with
other data samples (especially when performing \bp).

Praxis is a web application based on a client-server model. We implemented its
interface using Javascript with the help of D3~\cite{d3} and
ReactJS~\cite{react} libraries.  A separate analytics server carries out the
computations required by the projection computations. We implemented the server
in  Python, using the SciPy~\cite{scipy}, NumPy~\cite{numpy},
scikit-learn~\cite{scikit-learn} and CVXOPT~\cite{cvxopt} libraries.  We solve
the \bp generated quadratic optimization problems using CVXOPT~\cite{cvxopt}.

\section{Evaluation}
To evaluate our methods, we first conduct a user study with analysts and then
perform a computational model analysis assessing accuracy and scalability.    

\subsection{User Study}

We evaluate user experience with our techniques through a user study with
twelve data scientists. We have two goals. First, to assess the effectiveness
of our projection and visualization techniques in what-if analysis of
dimensionally reduced data.  Second, to understand how the use of the
techniques differs for changing task type and complexity.  

\noindent{\bf Participants:} We recruited twelve participants with at least two
years' experience in data science. Their areas of expertise included
healthcare analytics, genomics and machine learning. Participants ranged from
25 to 55 years old, and all had at least a Master's degree in science or
engineering. Ten participants  regularly applied  dimensionality reductions in
their data analysis, using Matlab, R and  Python. All participants were
familiar with using PCA, while six had used MDS before. Four participants cited
additional projection methods that they had previously used, including t-SNE
and autoencoder. 

\noindent{\bf Tasks and Data:} Participants were asked to perform the
following six high-level tasks using Praxis.  We used a tabular
dataset~\cite{Stahnke_2016,oecdDataset} containing eight socio-economic
development indices for thirty-four countries belonging to the Organisation for
Economic Co-operation and Development (OECD).  The dataset was a CSV file with
34 rows and nine columns, where one column represented country names.
Participants were free to use any combination of interactions and
visualizations to complete a given task. 

\begin{enumerate}
  \item[T1:] What four development indices contribute the most in determining the
    position of points in the projection plane?  Can you rank them based on
    their relevance? 
    
  \item[T2:] Can you explain why Portugal is in that specific position of the
    projection plane, distant from the other European countries? 

  \item [T3:] Suppose Chile has a near-term plan to attain a development level similar to Greece but could increase spending in only one of the
    development index areas. On which area would you advise the Chilean
    government to focus its resources? 

  \item [T4:] Consider the cluster formed by Turkey and Mexico. Compare it to
    the cluster formed by Asian countries.
  \item [T5:] Suppose Israel cannot increase its education spending  for the
    foreseeable future due to budgetary constraints.  Would it be reasonable
    for the country to attain a development level similar to Canada? 
  
  \item [T6:] Given that the Italian government would not allow
    WorkingLongHours to increase beyond the distribution mean, say which
    countries could be considered similar to Italy if it was able to improve
    its StudentSkills index value to 500.
\end{enumerate}
\noindent{\bf Procedure:} The study took place in the 
experimenter's office; one user at a time used Praxis running on 
the experimenter's laptop. The study had three stages. In the first, 
participants were briefed about the experiment
and  filled out a pre-experiment questionnaire eliciting information about
their experience in data science and use of dimensionality-reduction
techniques.  In the second stage, participants were introduced to the Praxis
interface and to our techniques via a training dataset.  Five minutes were
dedicated to showing each user how to perform  \fp, \bp and constraints
formulation.  Participants were then introduced  to a new dataset and asked to
complete  the six tasks above.  Task duration was manually timed and subject
responses were collected through think-aloud protocol. Participants had at most
two trials to perform each task; in the event of a failure, they moved on to
the next task. For each task we also recorded whether the task was completed
with or without the experimenter's help. In the last stage, participants were
asked to complete a post-experiment questionnaire to gather subjective
feedback.

\begin{table}
 \setlength{\tabcolsep}{3pt}
\centering
\begin{tabular}{*{8}{c}} 
 & \multicolumn{3}{c}{performance} & \multicolumn{4}{c}{techniques used} \\
                \cmidrule(lr){2-4}
                \cmidrule(lr){5-8}
Task & C & H & I & \emph{forward} & \pl & \emph{backward} & \emph{feasibility} \ \\ 
\midrule
T1  & 10  & 2  & 0  &  2  & 12  &  0  & 0  \\
T2  & 12  & 0  & 0  &  0  & 12  &  0  & 0  \\
T3  & 11  & 1  & 0  &  8  & 9   &  4  & 0  \\
T4  & 12  & 0  & 0  &  1  & 12  &  2  & 0  \\
T5  & 11  & 1  & 0  &  2  & 12  & 11  & 0  \\
T6  & 9   & 2  & 1  &  2  & 2   &  2  & 10 \\ 
\bottomrule
\end{tabular}
\caption{\normalfont Results of user study. Of a total of twelve participants, the table
  indicates for each task how many of them completed the task with no help (C),
  completed it with help (H) or did not complete the task (I). We also show the
  number of participants who used each of our proposed techniques to perform
single tasks. Note that both \pl and \fm can give users visual information
without requiring them to perform \fp or \bp (whereas, for instance, \fp is
intrinsically bound to \pl).\label{tab:results}}
\end{table}

\begin{figure}[h]
\centering
\includegraphics[width=\linewidth]{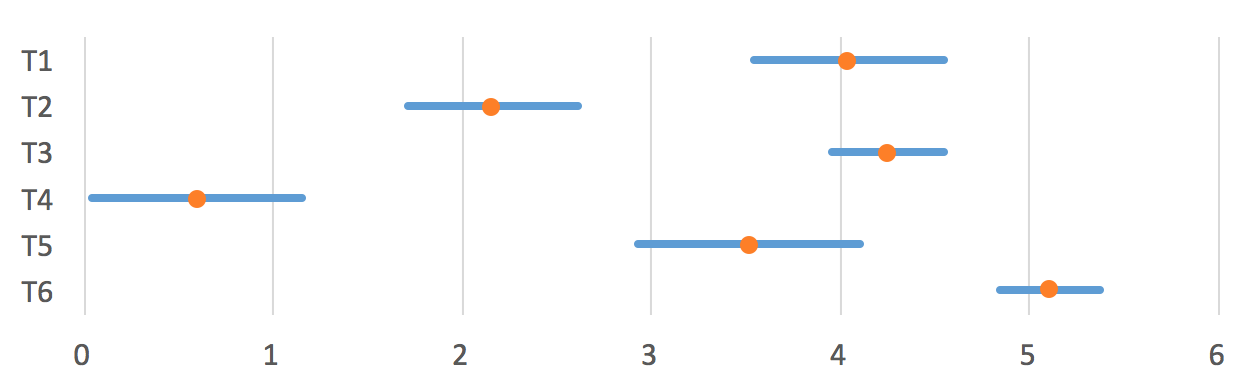}
\caption{\normalfont Task completion time. Average log time for participants to
complete the six assigned tasks.\label{fig:time_results}}
\end{figure}
\noindent{\bf Results and Discussion:} We adopt a task performance criterion
similar to that employed in \cite{Stahnke_2016}. For each task, we count the
number of participants who completed the task (C), completed the task with help
(H) or were unable to complete the task (I). Similarly, we also report the
frequency of the interaction (\fp and \bp)  and  visualization (\pl and \fm)
techniques used by users to complete each task.  We list these results in
Table~\ref{tab:results}.  Figure~\ref{fig:time_results} shows the average log
time spent on each task, inclusive of the cases in which the participant didn't
complete the task. All the completed tasks were performed in less than 30
seconds. Task completion times and their standard errors reflect the intrinsic
complexities of the tasks.  
 
Overall, \pl proved to be a simple yet powerful visualization technique for
exploring dimensionally reduced data as well as reasoning about the
dimensionality reduction. \Pl were used 59 times by participants over the
course of the six tasks performed. One participant mentioned he particularly
liked ``how prolines generate meaningful axes in a scatter plot where a clear
mapping to data dimensions is unclear,'' whereas another described prolines as
a ``great way to understand dimensionality reductions, especially for people
who used to treat them as a black box.'' The second most frequently used
technique was \bp, (used 19 times, followed  by \fp (15 times). Note that the
use of \fp always involves displaying \pl, which incorporates paths of \fp
locations for hypothetical feature values. Participants used \fp when they
wanted to interactively change the feature values and see precisely the
projection change of the corresponding data point. In particular, one
participant declared, ``I feel \bp is more natural to use and useful to see
which features correlate to each other, but I would prefer \fp for more precise
control over feature values.''  Also, despite its lower incidence of use, \fm
was employed by the participants when the task was sufficiently complex (T6). 

\begin{figure*}[t]
\centering
\includegraphics[width=0.9\linewidth]{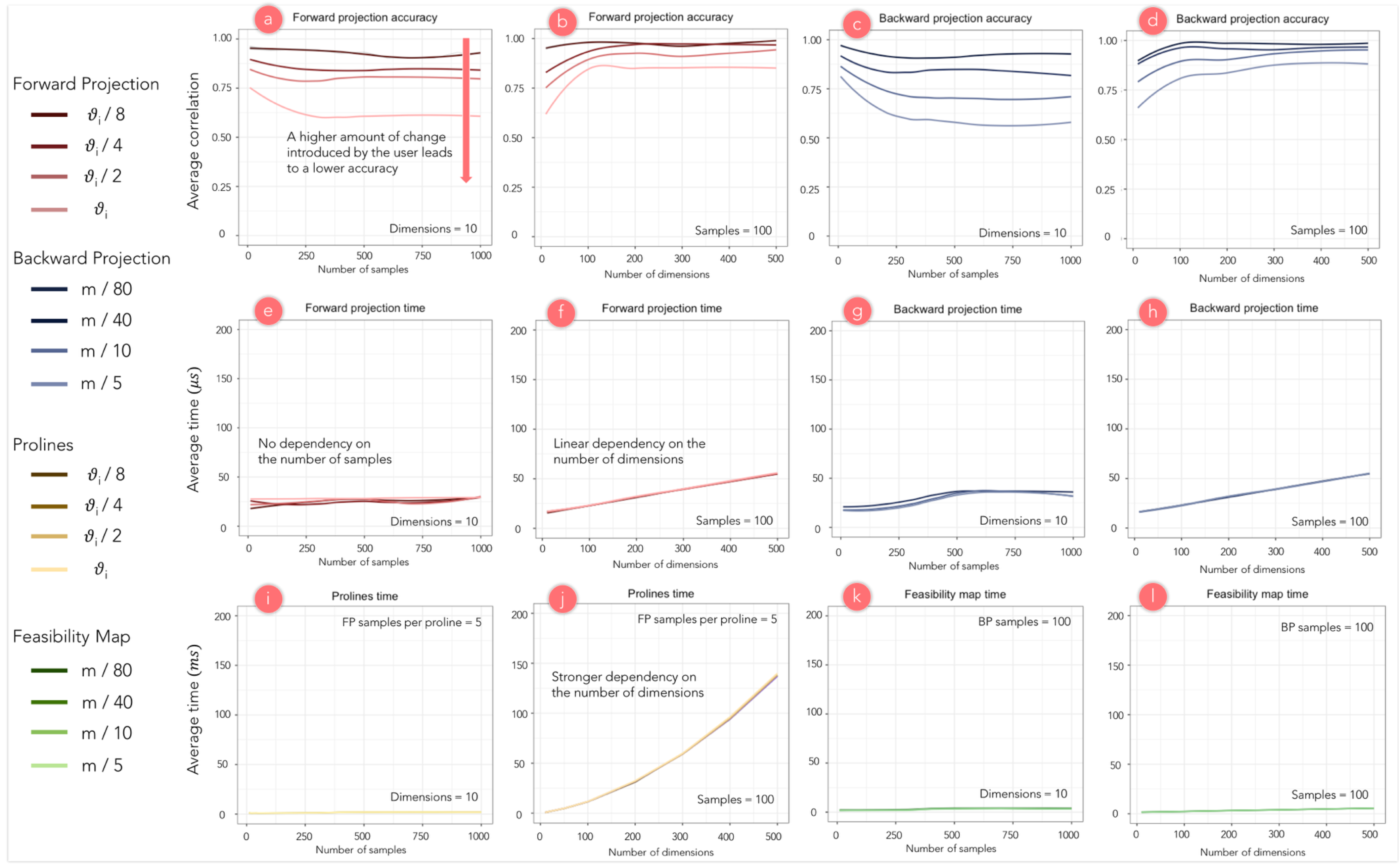}
\caption{\normalfont Accuracy and time performance results for PCA. We note that
the accuracy and time performance of \fp and \bp are mostly insensitive to the
number of samples and dimensions. Accuracy is instead tied to the amount of
change introduced by the user. The computational time for generating \pl shows
a linear dependence on the number of dimensions. Note that time is displayed in
microseconds in charts (e-h) and in milliseconds in charts (i-l).
\label{fig:analysis}}
\end{figure*}

\subsection{Model Analysis} 

Since they are intended for use in interactive applications, the computational
complexity of the proposed techniques needs to adhere to certain responsiveness
requirements. At the same time, forward and backward projection methods need to
be accurate enough at estimating changes in the dimensionally reduced space as
well as in the underlying multidimensional data so as not to lead the user to
false assumptions. We evaluate our proposed techniques for PCA in terms of time
and accuracy over varying number of samples and dimensions of the input dataset
and also over the amount of change introduced by the user (i.e. how much a
feature value is modified in the case of \fp, how much a data point is moved in
the case of \bp).

In our evaluation we iteratively perform \fp and unconstrained \bp on
automatically generated Gaussian random multivariate distributions, changing
either the number of data samples or the number of data dimensions and leaving
the other one fixed. We apply our techniques for each data point of the
original distribution. The \fp algorithm is then applied to each dimension with
an amount of change in $\left\{\sigma_i/8, \sigma_i/4, \sigma_i/2,
\sigma_i\right\}$, where $\sigma_i$ is the standard deviation for the current
feature. \Bp is performed in eight possible directions of movement (horizontal
and vertical axes plus diagonals), with an amount of change in $\left\{m/80,
m/40, m/20, m/10\right\}$, where $m$ is the width of the projection plane.
Accuracy and time performances are determined for each execution of the two
techniques and then averaged over all dimensions (directions), data samples and
test iterations.  All experimental results presented were generated on a
MacBook Pro, 2015 edition.

\noindent{\bf Time Performance:} Figure~\ref{fig:analysis} shows that the
execution time for both \fp (e,f) and \bp (g,h) is on the order of microseconds
and is not influenced by the number of samples nor by the amount of change;
charts i and h show a linear dependence on the number of dimensions that  does
not, however, significantly affect the time performance. Even when dealing with
larger datasets (e.g. $>$ 500 samples, $>$ 100 dimensions), both techniques are
suitable for interactive data analysis tools.  Figure~\ref{fig:analysis} also
shows the time performance of \pl (i,j) and \fm (k,l), respectively assuming
the computation of each proline with an average resolution of 5 \fp samples and
the generation of the \fm with a resolution of 100 \bp samples. In particular,
we note that the time to compute \pl depends linearly on the number of
dimensions.

\noindent{\bf Accuracy:} To assess the accuracy of our techniques, we
introduce a new similarity criterion for data-point neighborhoods in
dimensionally reduced spaces. For each execution of the algorithms on a data
point, we compute two sets of neighbors: 1) the $n$ closest neighbors in the
projection plane after performing \fp or after moving the data point with \bp,
and 2) the $n$ closest neighbors in the projection plane after performing the
dimensionality reduction on the multidimensional data, after it has been
modified through \fp or by \bp. Optimally, these two neighborhoods should
contain the same elements, which should have the same relative distance from
the data point on which the technique is performed.  We define a neighborhood
correlation index $c_n=c_e\times c_o$, where $c_e$ is the percentage of elements
that appear in both neighborhoods, whereas $c_o$ is the percentage of elements
whose distance from the data point considered remains in the same order. The
index varies between 0 and 1, with 1 corresponding to very similar
neighborhoods. 
Figure~\ref{fig:analysis} shows that the accuracy of both \fp (a,b) and \bp
(c,d) is mostly insensitive to the number of samples or dimensions. Instead, we
notice a strong dependence on the amount of change introduced by the user. This
shows that our proposed techniques are well suited for local changes in the
data, and greater user modifications could possibly alter the properties of the
dimensionality reduction.

\section{Interacting with Nonlinear Dimensionality Reductions}

We have so far demonstrated our interaction methods on PCA, a linear projection
method.  What about interacting with nonlinear dimensionality reductions?
There are out-of-sample extrapolation methods for many nonlinear
dimensionality-reduction techniques that make the extension of \fp with \pl
possible~\cite{Bengio_2004}. As for \bp, its computation will be
straightforward in certain cases (e.g. when an
autoencoder~\cite{hinton2006reducing} is used).  In general, however, some form
of constrained optimization specific to the dimensionality-reduction algorithm
will be needed. Nonetheless, it is highly desirable to develop general methods
that apply across dimensionality-reduction algorithms. 

Below we discuss an application of our techniques to an autoencoder-based
nonlinear dimensionality reduction.

\subsection{Autoencoder-based dimensionality reduction} 
An autoencoder is an artificial neural network model that can learn a
low-dimensional representation (or encoding) of data in an unsupervised
fashion~\cite{rumelhart1986learning}. Autoencoders that use multiple hidden
layers can discover nonlinear mappings between high-dimensional datasets and
their low-dimensional representations.  Unlike many other
dimensionality-reduction methods, an autoencoder gives mappings in both
directions between the data and low-dimensional
spaces~\cite{hinton2006reducing}, making it a natural candidate for application
of the interactions introduced here.  

\begin{figure} 
\centering
\includegraphics[width=\linewidth]{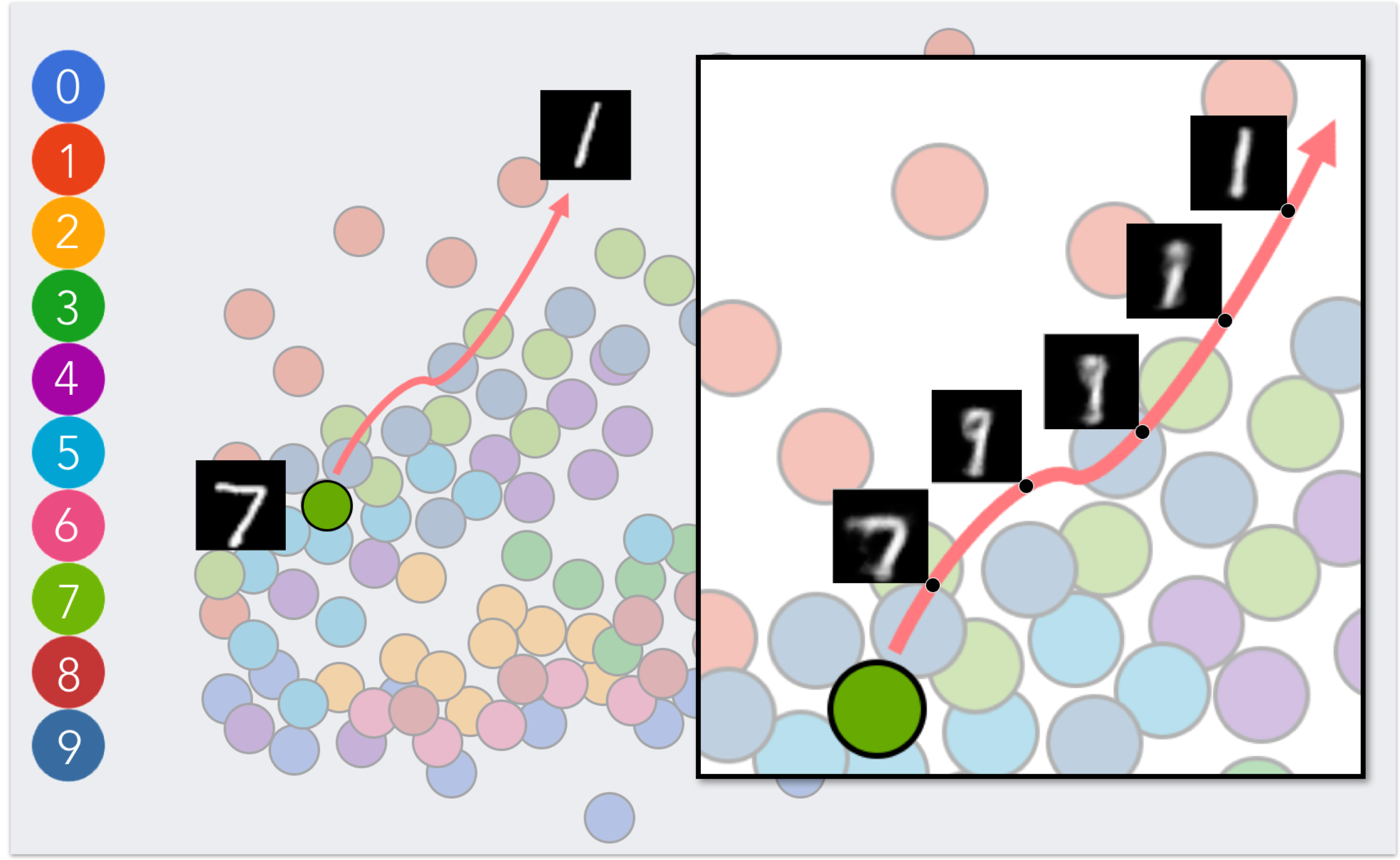} 
\caption{\normalfont Backward projection with autoencoder. A user explores a
  dimensionality reduction of handwritten digits from the MNIST
  database~\protect\cite{lecunmnist} using backward projection. The two-dimensional
  projection  is obtained with a deep autoencoder. Projected data points are colored 
  based on the digit represented.  By moving a node of the digit 7, back projection 
  enables the user to see how its feature values (pixels) are updated. In particular, 
  the user observes a smooth transition from 7 to 1, with 9 as an intermediate 
  representation.\label{fig:autoencoders}} 
\end{figure}

Figures~\ref{fig:autoencoders} shows how \bp can be applied in exploring  a
two-dimensional, autoencoder-based projection of handwritten digits from the
MNIST database~\cite{lecunmnist}. To this end, we first train an autoencoder
with seven fully connected layers using the 60,000-digit images from the
training set of the MNIST database. 

Each image had $28\text{ px} \times 28\text{ px} = 784\text{ features}$.  The
seven layers of the autoencoder had sizes 784, 128, 32, 2, 32, 128 and 784.
After training the network, we plot the encoded low-dimensional representations
of 100 digits from the MNIST test set as circular nodes in the plane. We color
each node by their associated digit. By performing \bp on a data point, it is
possible to observe the change in its feature values (pixel intensities) as the
point is moved around the projection plane. Our technique shows, in this case,
how one handwritten digit can gradually be transformed into another.

\begin{figure} \centering
  \includegraphics[width=\linewidth]{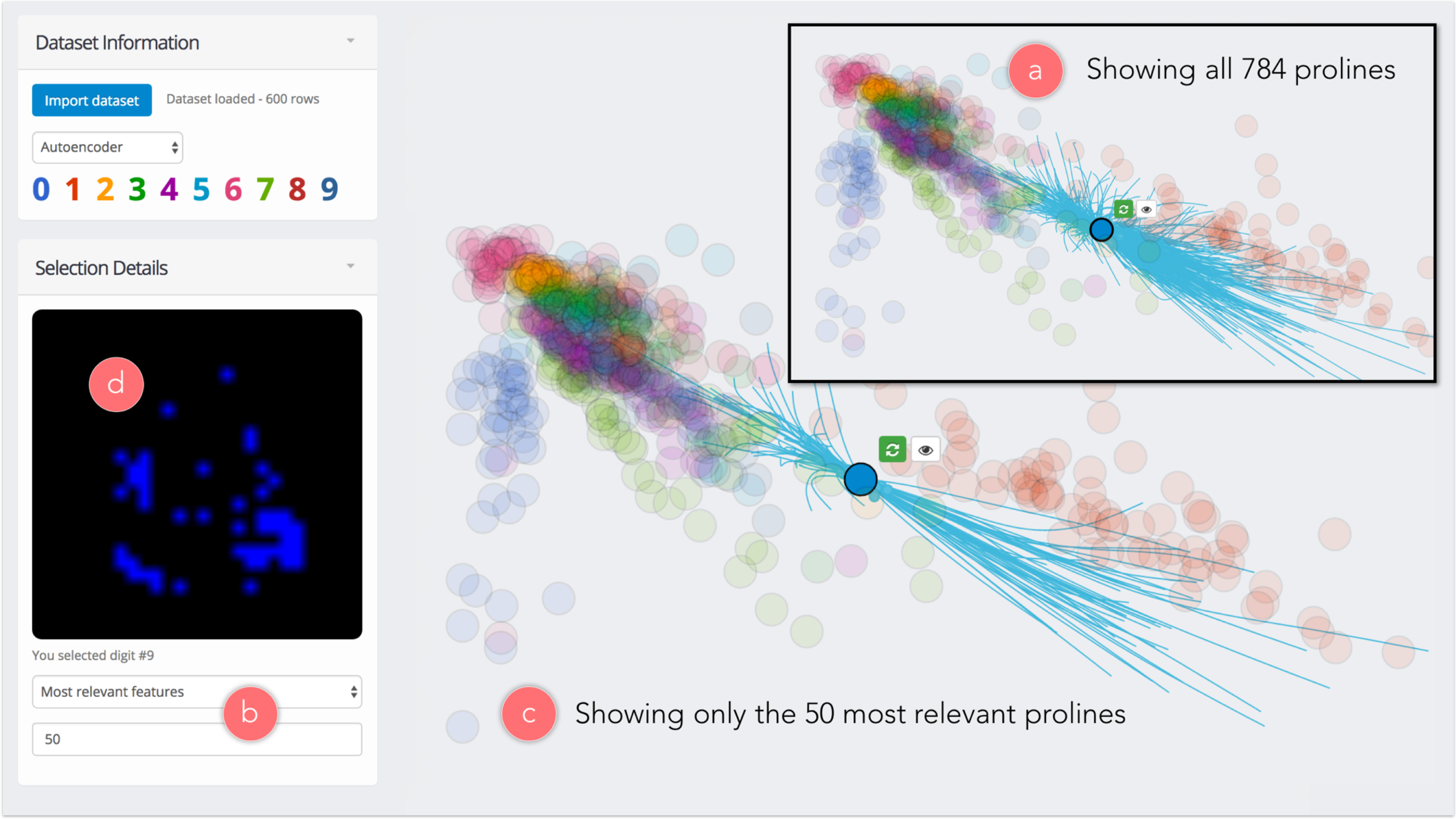}
  \caption{\normalfont Prolines (a) from an autoencoder-based projection of
    MNIST images~\protect\cite{lecunmnist}. A user can opt to show only the $n$
    most relevant prolines through Praxis' \textit{Selection Details} panel
    (b).  Fifty most relevant prolines (c) corresponding to the 50 pixels with
    the highest variability (d). In contrast to PCA, \pl in this case are not
    straight lines.  Depending on the user selection (b), the image in the
    \textit{Selection Details} panel can alternatively display: 1) the feature
    values (pixels) of the selected data point, 2) the difference map between
    the original pixels and their value after performing \fp or \bp, and 3) an
    interactive image for setting constraints on the values of each
    pixel.\label{fig:autoencoder_prolines} } 
  \end{figure}

Using the same neural network model, we integrate support for autoencoder-based
dimensionality reduction in our tool \textit{Praxis}. In particular,
Figure~\ref{fig:autoencoder_prolines} shows \pl are not straight lines for
non-linear dimensionality reduction methods.

\section{A Model for Dynamic Visualization Interactions}

The interaction techniques introduced here belong to a class of   interactions
that tightly couples data with its visual representation so that when users
interactively change one, they can observe a corresponding change in the other.
For example, through \fp, users  observe how the visual representation (2D
position) changes as they change the value of a dataset's attributes.
Conversely, users  can see how the data changes through \bp  as they change the
visual representation.  This class of interactions is  essential for realizing
dynamic visualizations (e.g.~\cite{Victor_2013,kondo2014dimpvis}) and we call
them {\em dynamic visualization interactions} or {\em dynamic interactions} for
short.

\begin{figure}[t]
\includegraphics[width=\linewidth]{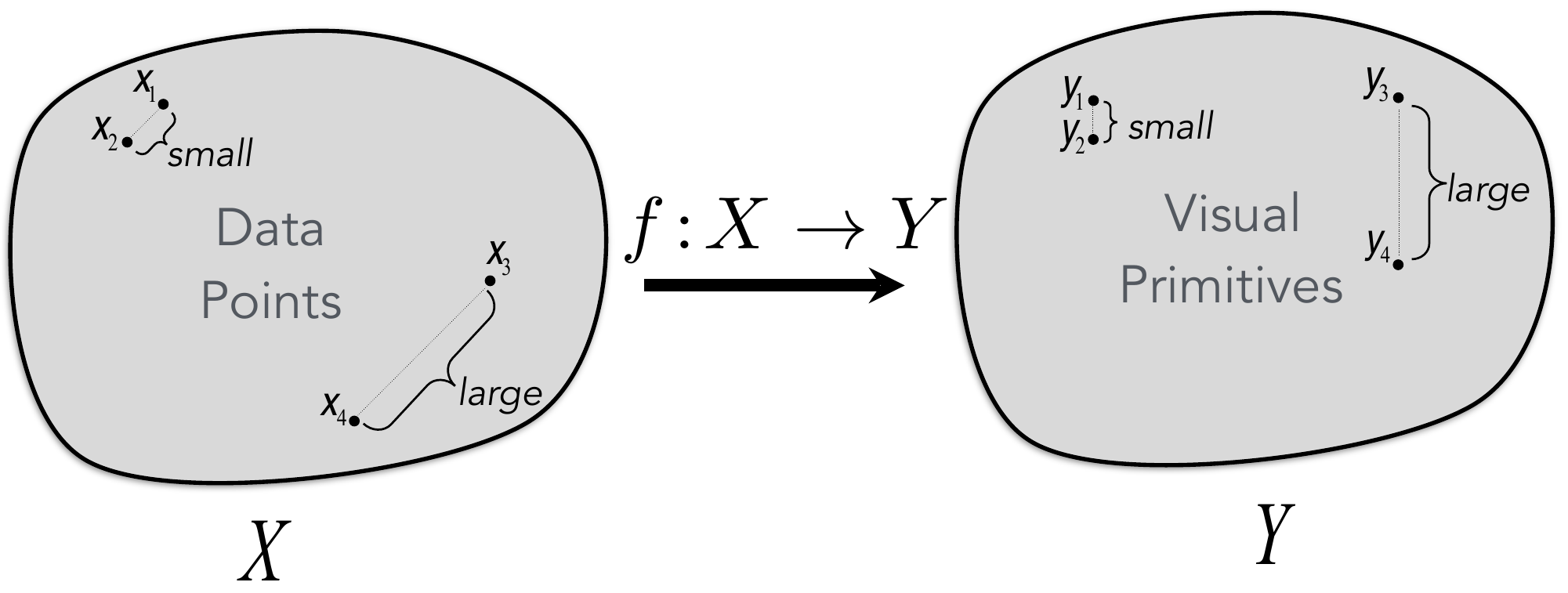}
\caption{\normalfont Visual embedding is a function that 
preserves structures in the data domain $X$ within the embedded 
perceptual space $Y$ (adapted from \protect\cite{Demiralp_2014}).\label{fig:ve}} 
\end{figure}

\begin{figure}[t]
\includegraphics[width=\linewidth]{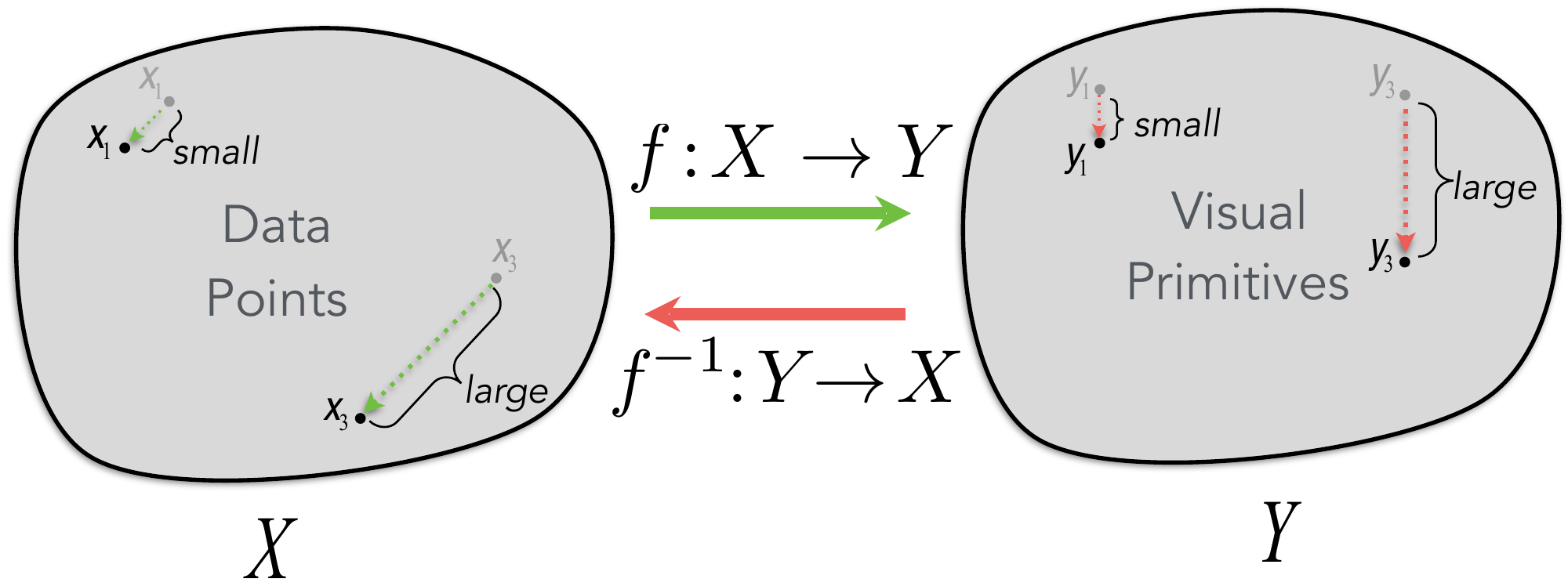}
\caption{
  \normalfont Bidirectionally coupling data and its visual
  representations. Visual embedding suggests that a change in the data should
  be reflected with a proportional change in its visual representation using
  $f$, the visual embedding function.  Conversely, change in a visual
  representation should be reflected by a proportional change in the
  corresponding data using $f^{-1}$.\label{fig:ve_extended}
} 
\end{figure}

We now look at dynamic interactions under the visual embedding
model~\cite{Demiralp_2014}. The visual embedding model  provides a functional
view of data visualization and posits that a good visualization is a structure-
or relation-preserving mapping from the data domain to the range (co-domain) of
visual encoding variables (Figure~\ref{fig:ve}). Visual embedding immediately
gives us criteria on which dynamic  interactions should be considered effective
(Figure~\ref{fig:ve_extended}): 1) a change in data (e.g., induced by user
through direct  manipulation) should cause a proportional change in its visual
representation and 2) a perceptual change in a visual encoding value (e.g., by
dragging nodes in a scatter plot or changing the height of a bar in a bar
chart) should be reflected by a change in data value that is proportional to
the perceived change.  However, to enable a dynamic interaction on a
visualization, we need to have access to both the visualization function $f$
and its inverse $f^{-1}$. The visual embedding model also  suggests why
implementing back  mapping to the data space  can be challenging. 

\begin{figure}[h]

\includegraphics[width=\linewidth]{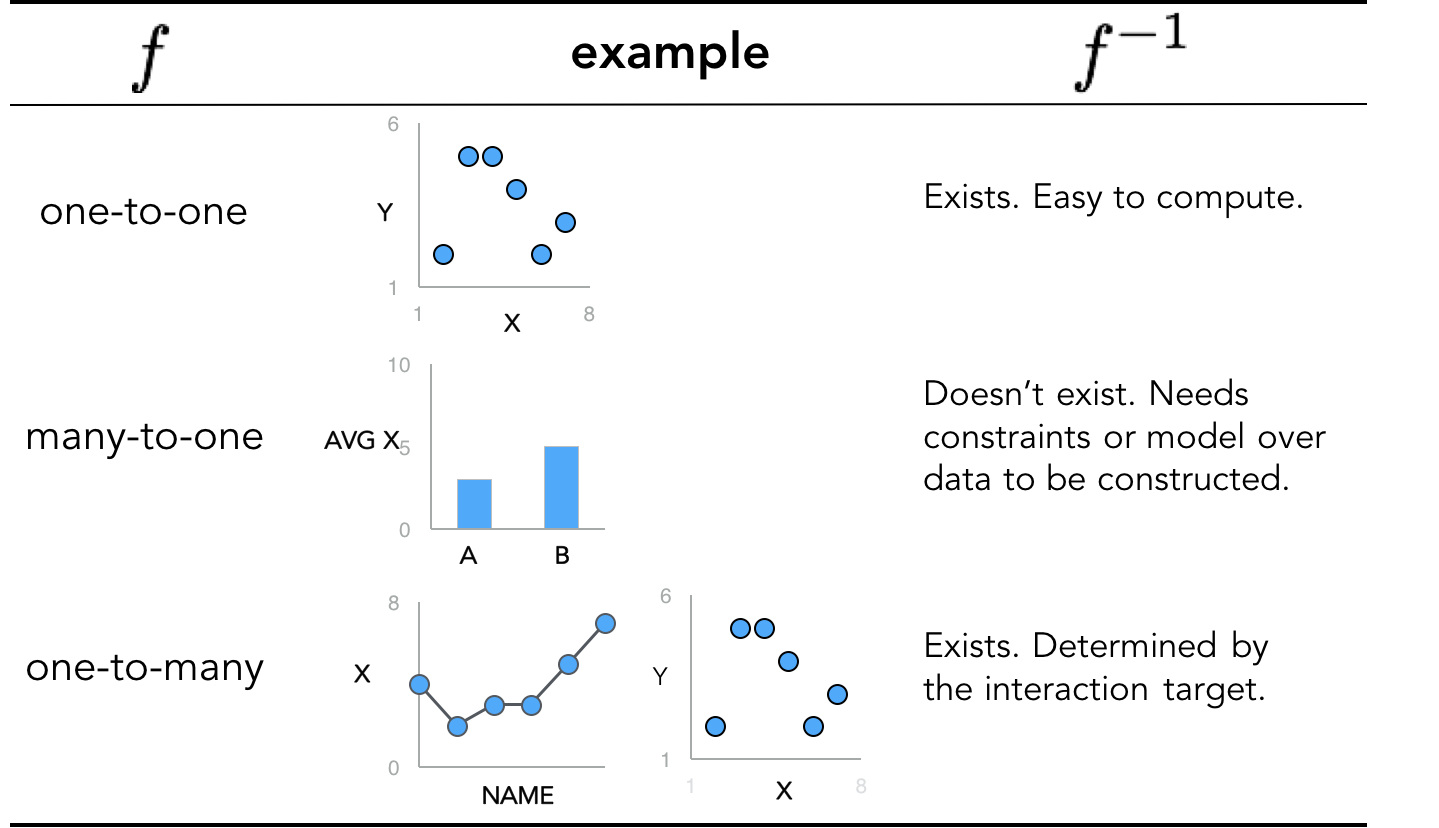}

\caption{\normalfont Visualization classes. Three basic forms of the visualization
  function $f$. Implementing  a dynamic interaction is clearly challenging when
  $f^{-1}$ does not exist.\label{fig:ve_classes} } 

\end{figure}

\begin{figure}[h]

\includegraphics[width=\linewidth]{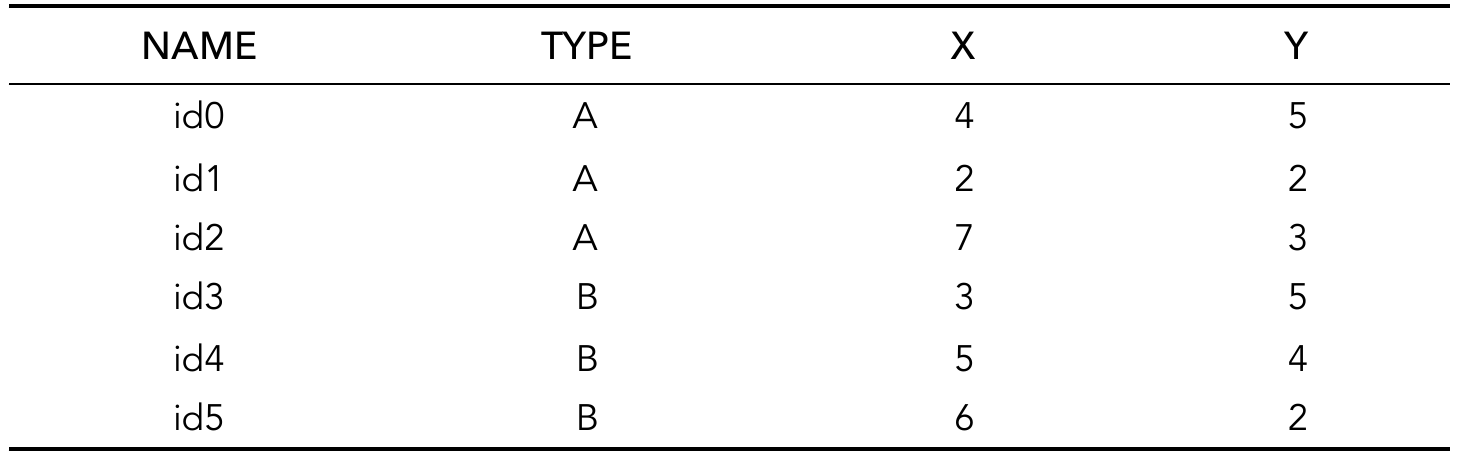}
\caption{ \normalfont Toy tabular dataset used for the three examples 
in Figure~\ref{fig:ve_classes}. \label{fig:ve_example_table}} 
\end{figure}

We consider three basic forms of the visualization function $f$ in
Figure~\ref{fig:ve_classes} through examples using a toy dataset in
Figure~\ref{fig:ve_example_table}. When the visualization function $f$ is
one-to-one, a dynamic interaction over $f$ is straightforward as $f^{-1}$
exists. When $f$ is one-to-many (still invertible but not necessarily a
proper function),  $f^{-1}$ exists and is determined by the target of
interaction. Consider the example in Figure~\ref{fig:ve_classes}. We visualize how
X values change for each NAME with a line chart. We also visualize the
correlation of X and Y with a scatter plot. If a user moves a point up or down
in the line chart, the corresponding change in X can be easily computed and the
scatter plot can be updated in a brush-and-link fashion. Essentially, the
one-to-many case can be seen as a collection of multiple one-to-one
visualizations. 

The most interesting case is when $f$ is many-to-one and hence not invertible.
A frequent source of such visualization  functions is summary data
aggregations, which are lossy. We can consider dimensionality reduction as a
form of aggregation. A simple example of a many-to-one visualization is the bar
chart (Figure~\ref{fig:ve_classes}) that shows the mean X for the data points
grouped by TYPE, A and B.  Now, in a dynamic interaction scenario, how should
we update the data values if a user changes the heights of the bars? Our \bp
solves a similar problem under a more complex visualization function, 
dimensionality reduction.  In
general, constructing a dynamic interaction over many-to-one visualization
functions would require imposing a set of assumptions over data  in the form
of, e.g., constraints or models. This presents a challenging yet important
future research direction. 

\section{Visual Analysis Is Like Doing Experiments} 

This paper introduces \fp, \bp, and the related visualizations, \pl and \fm, to
improve user experience in exploratory data analysis using DR.  We demonstrate
these techniques on PCA- and autoencoder-based DRs. We also contribute a new
tool, Praxis, implementing our techniques for DR-based data exploration. We
evaluate our work through a computational performance analysis, along with a
user study.  Our visual interactions are scalable, intuitive to use and
effective for DR-based exploratory data analysis. We situate our techniques in
a class of visualization  interactions at large that we discuss under the
visual embedding model.  

Data analysis is an iterative process in which analysts essentially run mental
experiments on data, asking questions, (re)forming and testing hypotheses.
Tukey and Wilk~\cite{Tukey_1966} were among the first in observing the
similarities between data analysis and doing experiments, listing eleven
similarities between  the two. In particular, one of these similarities states
that ``Interaction, feedback, trial and error are all essential; convenience is
dramatically helpful.'' In fact, data can be severely underutilized (e.g.,
\textit{dead}~\cite{Haas_2011,Victor_2013}) without what-if analysis.      

However, to perform data analysis as if we are running data experiments,
dynamic visual interactions that bidirectionally couple data and its visual
representations must be one of our tools. Our work here contributes to the
nascent toolkit needed for performing visual analysis in a similar way to
running experiments.

\section{Acknowledgments} 
The authors thank Paulo Freire for inspiring the name ``Praxis.''
\bibliographystyle{SIGCHI-Reference-Format}
\bibliography{paper}
\end{document}